\documentclass[twocolumn,showpacs,preprintnumbers,amsmath,amssymb]{revtex4}

\usepackage{graphicx}
\usepackage{dcolumn}
\usepackage{bm}

\begin{document}

\preprint{}

\title{Optical, vibrational, thermal, electrical, damage and phase-matching properties of lithium thioindate}

\author{S. Fossier, S. Sala\"{u}n, J. Mangin and O. Bidault}
\affiliation{%
Laboratoire de Physique de l'Universit\'{e} de Bourgogne (UMR-CNRS
5027), 9 avenue A. Savary, F-21078 Dijon, France.}%

\author{I. Th\'{e}not and J.-J. Zondy}
\email{jean-jacques.zondy@obspm.fr}
\affiliation{BNM-SYRTE, Observatoire de Paris (UMR-CNRS 8630), 61
avenue de l'Observatoire, F-75014
Paris, France.}%

\author{W. Chen}
 \affiliation {Laboratoire de Physicochimie de
 l'Atmosph\`{e}re (UMR-CNRS 8101), Universit\'{e} du Littoral C\^{o}te d'Opale, 145 Av. Maurice Schumann, F-59140 Dunkerque, France.}

\author{F. Rotermund}
\altaffiliation[Present address: ]{Department of Molecular Science
and Technology, Ajou University, 5 Wonchun-dong, Paldal-gu,
442-749 Suwon, Korea}
\author{V. Petrov}
\affiliation{Max-Born Institute for Nonlinear Optics and Ultrafast
Spectroscopy, 2A Max-Born St, D-12489 Berlin, Germany.}

\author{P. Petrov and J. Henningsen}
\affiliation{Danish Institute of Fundamental Metrology, B 307
Matematiktorvet, DK-2800, Lyngby, Denmark.}

\author{A. Yelisseyev, L. Isaenko and S. Lobanov}
\affiliation{Design and Technological Institute for Monocrystals
SB-RAS, 43 Russkaya St, RU-630058, Novosibirsk, Russia.}

\author{O. Balachninaite}
 \altaffiliation[Also at: ]{Laser Research Centre, Quantum Electronics Department, Vilnius University,
    LT-2040 Vilnius, Lithuania}
\author{G. Slekys}
 \altaffiliation[Also at: ]{Laser Research Centre, Quantum Electronics Department, Vilnius University,
    LT-2040 Vilnius, Lithuania}
\author{V. Sirutkaitis}
 \altaffiliation[Also at: ]{Laser Research Centre, Quantum Electronics Department, Vilnius University,
    LT-2040 Vilnius, Lithuania}
\affiliation{Altechna Co. Ltd, Ukmerges 41A-203, LT-2004, Vilnius,
Lithuania.}

\date{\today}

\begin{abstract}
Lithium thioindate (LiInS$_{2}$) is a new nonlinear chalcogenide
biaxial material transparent from 0.4 to 12 $\mu$m, that has been
successfully grown in large sizes and good optical quality. We
report on new physical properties that are relevant for laser and
nonlinear optics applications. With respect to AgGaS(e)$_2$
ternary chalcopyrite materials, LiInS$_{2}$ displays a
nearly-isotropic thermal expansion behavior, a 5-times larger
thermal conductivity associated with high optical damage
thresholds, and an extremely low intensity-dependent absorption
allowing direct high-power downconversion from the near-IR to the
deep mid-IR. Continuous-wave difference-frequency generation
(5-11$\,\mu$m) of Ti:sapphire laser sources is reported for the
first time.
\end{abstract}

\maketitle

\section{\label{sec: introduction}Introduction}

The intensive search for new nonlinear materials for the
generation of coherent tunable radiation in the mid-IR
($2-20\,\mu$m), a spectral range of importance for molecular
spectroscopy, atmospheric sensing and various opto-electronic
devices, remains a continuing challenge. In contrast with the
oxides (with IR transmission range not exceeding $\sim 5\,\mu$m)
that attracted much attention lately, such deep mid-IR materials
cannot be grown by well mastered hydrothermal, flux or Czochralski
methods. Instead the more complex Bridgman-Stockbarger growth
technique in sealed (high-pressure) ampoules, with volatile and
chemically reactive starting components, is the only method used
to produce large size single domain crystals, and this certainly
hampered their development all the more that special post-growth
treatments are needed to restore stoichiometry and improve their
optical quality.

To date, only few suitable nonlinear crystals combining an
extended transparency from the visible/near-IR to the deep mid-IR
and large enough birefringence to allow phase-matching over the
whole transparency range are available. The majority of these
compounds belong to the ternary chalcogenide semi-conductors of
the A$^{\text{I}}$B$^{\text{III}}$C$_{2}^{\text{VI}}$ family,
where A=Cu, Ag; B=Al, Ga, In; C=S, Se, Te~\cite{bhar}. Except for
AgInS$_2$, which can exist in both the tetragonal and orthorhombic
phases~\cite{shay}, these crystallize in the chalcopyrite
(CuFeS$_{2}$) structure with $I\bar{4}2d$ space group and
tetragonal symmetry (class) $\bar{4}2m$ but only few of them
possess sufficient birefringence. Most of the commercially
available nonlinear crystals (AgGaS$_{2}$ or AGS, AgGaSe$_{2}$ or
AGSe, ZnGeP$_{2}$ or ZGP, GaSe, CdSe) or more confidential ones
(HgGa$_{2}$S$_{4}$ or HGS, Ag$_{3}$AsS$_{3}$ or proustite,
Tl$_{3}$AsSe$_{3}$ or TAS) have their advantages and
drawbacks~\cite{dmitriev}. Despite their higher nonlinearity
compared with the oxides, one limitation stems from their low
energy bandgap, which prevents the use of pump lasers below 1
$\mu$m due to severe linear and two-photon absorption and
associated thermal effects. ZGP has a high nonlinearity (72 pm/V)
and thermal conductivity but requires pump wavelength above 2
$\mu$m due to its residual absorption in the near-IR. The noble
metal compounds (AGS, AGSe) have the lowest residual absorption,
but their poor thermal conductivity due to low-energy phonon
spectrum limits their performance, especially in cw
applications~\cite{stoll,lee,douillet1,douillet2}. Furthermore,
their thermal expansion is anisotropic along the direction
parallel and perpendicular to the optic axis, which is a source of
thermo-mechanical stresses. CdSe's transparency extends up to
$18\mu$m but its low birefringence and nonlinearity limit its
phase-matching capability. GaSe is a soft, cleaving compound that
cannot be cut at directions different from the optical axis,
further its large birefringence results in severe walk-off
limitations.

Two new materials belonging to the
A$^{\text{I}}$B$^{\text{III}}$C$_{2}^{\text{VI}}$ chalcogenide
family, where the metal cations are replaced with the lighter
alkali metal (A=Li), can now be added to this limited list:
LiInS$_{2}$ or lithium thioindate (LIS) and LiInSe$_{2}$ or
lithium selenoindate (LISe). They were shown to crystallize with
the $\beta$-NaFeO$_{2}$ structure (orthorhombic Pna2$_{1}\,\equiv$
$C_{2v}^9$ symmetry, point group $mm2$), which is a distorted
superstructure of the wurtzite type, as early as
1965~\cite{hoppe}. Hence, contrary to the other mid-IR compounds,
these new materials are biaxial and isostructural to, e.g.,
KTiOPO$_{4}$ (KTP). The linear and nonlinear properties of LIS
were briefly studied by Boyd and co-workers in the early 70's,
using small and poor quality samples~\cite{boyd}, but very few
works (mainly focused on its optical bandgap characterization and
photoluminescence
spectra~\cite{kamijoh,kamijoh2,nozaki,kuriyama1,kuriyama-kt,kuriyama2})
have followed since until recently due to problems associated to
its reliable growth in larger sizes allowing to design good
optical quality single domain elements for practical
applications~\cite{golovei}. Thin films of LiInS$_2$ on glass and
silicon substrates were also characterized~\cite{saitoh}. The
growth problems, associated with the high chemical activity of Li,
were solved recently~\cite{isaenko-vasilyeva} and large crystals
(up to 20 mm in length) of high optical quality became available
for practical nonlinear optical
applications~\cite{assl,knippels,rotermund}. To our knowledge,
there is only one other group which has succeeded previously in
synthesizing this material with comparable size~\cite{schumann},
but this work was also stopped recently. While only
non-phase-matched second-harmonic generation (SHG) was used by
Boyd and co-workers to determine the nonlinear properties of
LIS~\cite{boyd}, the first phase-matched SHG experiment using the
tunable output of a pulsed LiNbO$_{3}$-OPO near 2.5 $\mu$m was
reported only recently~\cite{assl}. The range of fundamental
wavelengths was extended to $6\,\mu$m using the radiation from a
free-electron laser and ZGP-frequency doubler~\cite{knippels} and
rough estimation of the effective nonlinear coefficients was
extracted from those measurements. Direct parametric
down-conversion in the range $4.8-9\,\mu$m, using a femtosecond
(200 fs) Ti:sapphire amplifier as the pump source, was
simultaneously reported with an upper-bound value of the
two-photon absorption (TPA) at 800 nm of $\beta$=0.04 cm
/GW~\cite{rotermund}, owing to the high bandgap energy of
LIS~\cite{kuriyama1,kuriyama-kt,schumann,eifler}. Such a value is
about 100 times less than the value for AGS or 1000 times less
than for AGSe at 1.32 $\mu$m~\cite{pearl}.

These preliminary experiments allowed to assess the potential of
LIS over the existing mid-IR materials, and justify the need for a
complete characterization of its optical properties. In this
paper, we report new results from an intensive investigation
campaign involving a network of laboratories in Europe, on the
characterization of the main properties relevant for laser
applications: structure and transmission (Section~\ref{sec:
growth}), specific heat, thermal expansion and thermo-optic
coefficients (Section~\ref{sec: thermal}), accurate lattice phonon
spectra (Section~\ref{sec: vibrational}), electro-optic and
piezo-electric constants (Section~\ref{sec: electro-piezo}),
linear optical and thermo-optic dispersions (Section~\ref{sec:
linear}), phase-matching predictions (Section~\ref{sec:
phase-matching}), nonlinear coefficients (Section~\ref{sec:
meas-deff}), parametric down-conversion (Section~\ref{sec:
down-conv}) and optical damage thresholds (Section~\ref{sec:
damage}). The results presented here show that LIS is superior to
the existing chalcopyrites in terms of high-power deep mid-IR
down-conversion applications pumped directly from the near IR. The
preliminary work on LISe has just started and will not be reported
here~\cite{isaenko-petrov}.

\section{\label{sec: growth}Composition, structure and transmission}

Single-domain crystals of LIS are grown by the
Bridgman-Stockbarger technique in a vertical setup with
counter-pressure on seeds along (001) and (010)
directions~\cite{hoppe,boyd,kamijoh,isaenko-vasilyeva,kovach,kish84,bruckner}.
The directed crystallization is performed from a melt of elemental
Li, In and S. In the melt zone the temperature is maintained at
$\sim 1100\,$K and decreases to $\sim 900\,$K in the growth zone.
The charged ampoules are moved in an optimal thermal gradient of
$10-15\,$K/cm at a rate of $\sim 10\,$mm/day.
 This technique allows the growth of crystal
ingots with diameters up to $20\,$mm and lengths up to
50mm~\cite{isaenko-vasilyeva}. Although sometimes colorless to a
great extent, the as-grown ingots are usually milky because of
small inclusions of various phases and it is necessary to anneal
them in Li$_{2}$S or S$_{2}$ vapor at temperature close to the
melting point ($T_{\mbox{\footnotesize{melt}}}\sim 1000^{\circ}$C)
~\cite{isaenko-vasilyeva}. The crystal coloration after the
thermal post-growth treatment changes then from almost colorless
(or slightly yellow) for as-grown, to salmon-rose tinge depending
on the growth and post-growth treatment conditions and chemical
composition. Optical elements with aperture up to $1\,$cm$^{2}$
and length up to $15\,$mm could be prepared for optical
measurements or for nonlinear frequency conversion.
  \begin{figure}[t]
    \includegraphics [width=6cm]{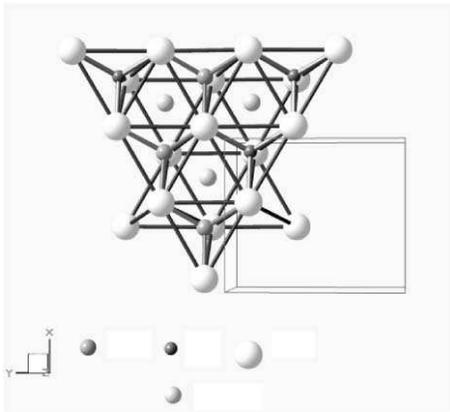}
    \caption{\label{fig: lis-struc} Fragment of the orthorhombic unit cell structure (the box frame gives the orientation of the
    unit cell). The 3 empty areas materialized with light grey spheres indicate the location of octapores.}
  \end{figure}

Precise chemical analysis - using differential dissolution
techniques (DDT) and inductively-coupled plasma (ICP) method -
showed that the composition of colorless (or yellow tinge) and
rose samples is Li$_{0.98}$In$_{1.03}$S$_{2.00}$
(Li$_{0.96}$In$_{1.06}$S$_{1.98}$) and
Li$_{0.86}$In$_{1.04}$S$_{1.98}$
respectively~\cite{isaenko-vasilyeva}. Lithium deficit
($N_{\mbox{\footnotesize{Li}}}:N_{\mbox{\footnotesize{In}}}<1$) is
observed for all samples, the annealed rose ones being
characterized by a higher sulfur content
($N_{\mbox{\footnotesize{Li+In}}}:N_{\mbox{\footnotesize{S}}}<1$).
The band-structure nature of LIS has been confirmed to be of the
\emph{direct} type (direct transitions between parabolic
bands)~\cite{kuriyama1,kuriyama-kt}. The forbidden band gap values
estimated by us ($E_{g}=3.72$ and 3.57 eV at 80K and 300K,
respectively, which are in good agreement with previous
measurements~\cite{kuriyama1,kuriyama-kt,schumann,eifler}), the
position of the long-wave edge of transparency ($13\mu$m at an
absorption of 6 cm$^{-1}$) and the crystal symmetry group were
found identical for all crystal colorations although small
variations in lattice parameters due to point defects were also
observed~\cite{zakopane,zakospie}.

A detailed and general structural analysis of
A$^{\text{I}}$B$^{\text{III}}$C$_{2}^{\text{VI}}$ compounds, where
A$^{\text{I}}$ is an alkali metal such as Li, has been given by
Kish \emph{et al}~\cite{kish87}. A fragment of the structure of
LIS is shown in Fig.~\ref{fig: lis-struc} (see Figs. 2.1.1-2.1.2
in Ref.~\cite{bruckner} for the full unit cell representation). It
is formed by LiS$_{4}$ and InS$_{4}$ tetrahedrons and the S$^{2-}$
ions are arranged in hexagonal packing with tetragonal and
octahedral cavities (tetra- and octapores)~\cite{kish87}. Compared
to the denser chalcopyrite structure of AGS or
AGSe~\cite{feigelson}, the $\beta$-NaFeO$_{2}$ structure of
Li-chalcogenides allows the presence of empty cavities within the
unit cell volume. The less dense $\beta$-NaFeO$_{2}$ structure
allows in particular doping with rare-earth (RE) active ions which
can be hosted inside the octapores. Such a preliminary doping
experiment with Nd$^{3+}$ ions was previously reported with the
aim of investigating RE:LIS potential as a self down-converting
material~\cite{ico}. The values of the orthorhombic lattice
parameters are $a=6.874(1)\mbox{\AA}$, $b=8.033(2)\mbox{\AA}$,
$c=6.462(1)\mbox{\AA}$ for as-grown colorless LIS, and
$a=6.890(1)\mbox{\AA}$, $b=8.053(1)\mbox{\AA}$,
$c=6.478(1)\mbox{\AA}$ for slightly yellowish samples. They are
slightly lower than the values reported by Hoppe~\cite{hoppe} and
Boyd \emph{et al}~\cite{boyd} or Kamijoh \emph{et
al}~\cite{kamijoh,kamijoh2} and Kish \emph{et al}~\cite{kish85},
but most closely to the last one. They modify to
$a=6.896(1)\mbox{\AA}$, $b=8.058(2)\mbox{\AA}$,
$c=6.484(4)\mbox{\AA}$ for rose annealed samples. Concomitantly,
the density changes from $\rho=3.52$g/cm$^{3}$ to
$\rho=3.44$g/cm$^ {3}$. We have chosen $c<a<b$ where $c$ is the
polar two-fold axis which coincides with the selection of Boyd
\emph{et al}~\cite{boyd}. This convention satisfies the
recommendation of Roberts~\cite{roberts} for unique designation of
the orthorhombic class $mm2$. Note that in the following, in
contrast to Roberts~\cite{roberts} and the ANSI/IEEE
Standard~\cite{ire-standard}, we do not introduce an additional
frame for reporting the nonlinear tensor properties but for
simplification and in accordance with the tradition use the
crystallographic $abc$ frame for reporting the nonlinear optical
susceptibility (see Section~\ref{sec: phase-matching}). Also,
instead of small case letters $xyz$, as suggested by
Roberts~\cite{roberts}, we use the capital letters $XYZ$ for
designation of the principal optical axes such that $n_X<n_Y<n_Z$.
Since for LIS one has $n_{b}<n_{a}<n_{c}$ the principal frame
assignment will be $X\leftrightarrow b$, $Y\leftrightarrow a$ and
$Z\leftrightarrow c$.

  \begin{figure}[t]
    \includegraphics[width=7cm]{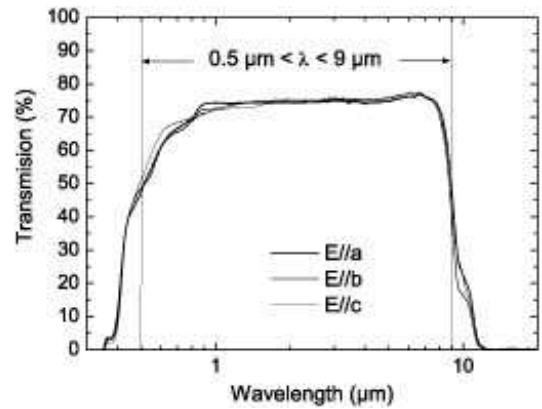}
    \caption{\label{fig: ftir} Polarized transmission spectra of an annealed LIS sample (thickness 5 mm), not corrected for the
    Fresnel loss.}
  \end{figure}

Fig.~\ref{fig: ftir} shows the polarized transmission spectra of a
thick ($L=5\,$mm) annealed optical element. Depending on the
growth and annealing conditions, some absorption bands due to the
various point defects - in particular excess sulfur ions occupying
interstitial sites - may limit the transmission window in the
visible side. These visible bands, responsible for the rose
coloration of the annealed samples, restrict the range of
high-transmission ($T\sim70\%$) to $0.8 - 8.1\mu$m. The
post-growth thermal treatment, however, improves considerably the
transparency within the above window. The absorption coefficient
$\alpha(\lambda)$ within the full transparency window deduced from
the spectra amounts to 0.1 - 0.15 cm$^{-1}$. These values are
overestimated, since from direct laser line transmission
experiments using various laser sources in the range
$0.78-2.53\,\mu$m and several samples the average residual
absorption of a typical annealed sample was found to be lower. At
$2.53\,\mu$m, e.g., the total loss coefficient (absorption and
scattering) estimated amounts to $0.05\,$cm$^{-1}$. At $780\,$nm
the loss coefficient measured with a diode laser source ranged
between 0.05 and 0.2 cm$^{-1}$ depending on the strength of the
weak point-defects related bands~\cite{isaenko-vasilyeva} seen
also in Fig.~\ref{fig: ftir}.

\section{\label{sec: thermal} Thermal and thermo-optic properties}

The thermally-related properties of a new nonlinear material are
crucial in assessing its potential in real nonlinear conversion
devices pumped by high power cw or pulsed lasers. The performance
of devices based on mid-IR chalcogenides is often limited by
deleterious thermal effects (lensing,
expansion)~\cite{stoll,douillet2,pearl}. The thermal figure of
merit of a material can be defined by the quantity
$\eta=(\text{d}n/\text{d}T)/K$ where $\text{d}n/\text{d}T$ is the
thermo-optic index variation and $K$ is the thermal conductivity.
The larger this quantity the more severe are the thermal
limitations. A second thermal quantity, $\tau_D=w_0C_p/4K$ where
$C_p$ is the mass specific heat of the material and $w_0$ is the
beam radius, gives the characteristic time of heat diffusion
outside the pumped volume. The larger this diffusion time, the
stronger the thermal lensing effects~\cite{douillet2}. It is hence
important to know $C_p, K$ and $\text{d}n/\text{d}T$. In addition,
the thermo-optic coefficients are needed to predict phase-matching
performance in $\chi^{(2)}$ interactions. Let us point out that no
data has been published on the thermal properties of single domain
crystalline LIS.

\subsection{\label{subsec: specheat} Specific heat at constant pressure}
Using Al$_2$O$_3$ as a reference we estimated $C_p(T=300K)=92.9
\pm 1.1\,$J/mol/K for crystalline rose colour LIS (57.91\,mg)
which is only slightly higher than what has been previously
measured for polycrystalline LIS (90\,J/mol/K) in
Ref.~\cite{kuhn}. The temperature dependence in the
[-160$^\circ$C; 210$^\circ$C] range could be fitted by
\begin{equation}
    C_p(T)=114.02-5708\times T^{-1}-187540\times T^{-2}
\end{equation}
where $T$ is in Kelvin. The general theoretical expression of
$C_{p}\,(T)$ for anharmonic solids is
\begin{equation} \label{eq: cpt}
    C_{p}\,(T)=12R\left[F(x_{D})+\sum_{k=1}^{N}c_{k}T^{k}\right]
\end{equation}
where $R$ is the molar gas constant, $x_{D}=T_{D}/T$ ($T_{D}$ is
the Debye temperature) and $F(x_{D})$ is the Debye function
describing the temperature dependence in the harmonic lattice
vibration approximation,
\begin{equation} \label{eq: fxd}
    F_{}(x_{D})=\frac{3}{x_{D}^{3}}\int_{0}^{x_{D}}\frac{x^{4}e^{x}dx}{(e^{x}-1)^{2}}.
\end{equation}
The value of $N$ and the absolute magnitudes of the coefficients
$c_{k}$ in the sum expansion in (\ref{eq: cpt}) can be considered
as a measure of the degree of lattice anharmonicity, while the
sign of $c_{k}$ is essentially determined by the shape of the
interatomic potential. In Eq.~(\ref{eq: cpt}), the leading
coefficients $c_{1}$ and $c_{3}$ were found negative for all
A$^{\text{I}}$B$^{\text{III}}$C$_{2}^{\text{VI}}$ chalcogenide
compounds, and the magnitude of $c_{k}$ was found $\sim10$ times
higher and independent of the anion C$^{2-}$ for Li-compounds.
This means that the anharmonic contribution of the lattice
potential energy is much stronger in LiInC$_{2}^{\text{VI}}$
compounds ($N=4$ in Eq.~\ref{eq: cpt}) than in
A$^{\text{I}}$B$^{\text{III}}$C$_{2}^{\text{VI}}$ chalcopyrite
compounds ($N=3$). In LIS, this is related to the specific nature
of the Li-S bond, which is about twice weaker than the In-S
bond~\cite{sobota}. As a consequence, the temperature dependence
of $C_{p}$(LIS) is much weaker than for AgInS$_{2}$ and
AGSe~\cite{neumann}, and the resulting $C_{p}(T)$ values for LIS
are the lowest of all chalcogenides. By a calorimetric absorptance
($A$) measurement the absorption coefficient $\alpha$ can be
estimated from $A=mC_p\Delta T/P\Delta t=1-\exp(-\alpha L)$ where
$m$ is the mass in mol, $\Delta T$ is the temperature rise, $P$ is
the laser power and $\Delta t$ is the irradiation time. We
irradiated at 1064\,nm an annealed LIS sample of dimension
$5\times 5\times 5\,$mm$^3$ cut at 28$^\circ$ from the $c$
crystallographic axis for light propagation in the a-c plane (mass
$m=0.438\,\text{g} =2.356\times 10^{-3}\,$mol) and obtained an
absorption coefficient of $\alpha=0.037\,$cm$^{-1}$ (compare
Section~\ref{sec: growth}).

\subsection{\label{subsec: thermal-expansion} Thermal expansion}

The principal thermal expansion and thermo-optic coefficients were
determined by using the absolute interferometric dilatometer and
the experimental procedure described in Ref.~\cite{dilatometer}. A
two-beam modified Mach-Zehnder interferometer arrangement is
employed for dilatation measurements, while thermo-optic
coefficients are obtained using the same set-up by recording
changes in optical thickness of a sample acting as a thermal
scanning Fabry-Perot interferometer, with the natural reflectivity
of the two opposite parallel facets. Three parallelepipedic
samples of dimension $5\times5\times8$ mm$^{3}$ cut along the
crystallographic axes $a$, $b$ and $c$ ($L=8\,$mm) were used for
these measurements. The samples were subjected to linear
temperature ramps of $0.2^{\circ}$C/min in the range
$-20^{\circ}$C - $+100^{\circ}$C.
  \begin{figure}[t]
    \includegraphics[width=7cm]{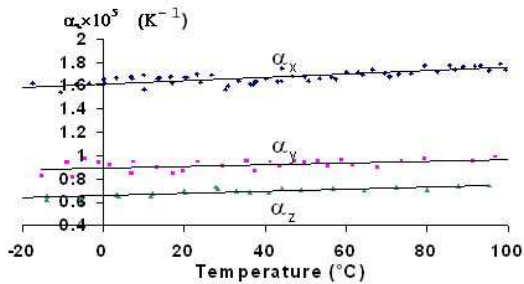}
    \caption{\label{fig: thermal-expansion} Thermal expansion coefficients of LIS along the axes $X\equiv b, Y\equiv
    a, Z\equiv c$ from $-20^{\circ}$C to $+100^{\circ}$C.}
  \end{figure}

In a given temperature interval $dT$ the linear thermal expansion
coefficients  are defined as
\begin{equation} \label{eq: alphai}
    \alpha_{i}=\frac{1}{L_{i}}\frac{dL_{i}}{dT}
\end{equation}
where $L_{i}$ is the sample length along the direction $i$, with
$i=a,b,c$. To be consistent with our assignment of the principal
optical axes $X,Y,Z$ in Section~\ref{sec: linear}, we shall once
and for all make the substitutions $X\equiv b, Y\equiv a, Z\equiv
c$ in the following.

For the measurement of the thermal expansion the light source of
the absolute interferometric dilatometer was a frequency
stabilized 2-mW He-Ne laser.  A plot of the experimental data is
reported in Fig.~\ref{fig: thermal-expansion}. A linear fit of the
data gives the following temperature dependence of the linear
thermal expansion coefficients,
\begin{eqnarray}
    \alpha_{X}=1.61\times 10^{-5}+1.4\times 10^{-8}T, \label{eq: alphax}\\
    \alpha_{Y}=0.89\times 10^{-5}+0.7\times 10^{-8}T, \label{eq: alphay}\\
    \alpha_{Z}=0.66\times 10^{-5}+0.9\times 10^{-8}T, \label{eq: alphaz}
\end{eqnarray}
where $T$ is the temperature ($^{\circ}$C). The deviation of
experimental values from these fits are less than
$10^{-6}\,$K$^{-1}$. At $T=20^{\circ}$C we have
$\alpha_{X}=1.64\times10^{-5}\,$K$^{-1}$,
$\alpha_{Y}=0.90\times10^{-5}\,$K$^{-1}$ and
$\alpha_{Z}=0.68\times10^{-5}\,$K$^{-1}$, respectively. Such
values are of the same order of magnitude (or even less) than
those for AGS ($\alpha_{\parallel c}=1.25\times10^{-5}\,$K$^{-1}$,
$\alpha_{\perp c}=-1.32\times10^{-5}\,$K$^{-1}$)~\cite{korczak}
and AGSe ($\alpha_{\parallel c}=1.68\times10^{-5}\,$K$^{-1}$,
$\alpha_{\perp c}=-0.78\times10^{-5}\,$K$^{-1}$)~\cite{iseler}.

However, the main difference with those chalcopyrites is that LIS
does not display their anomalous thermal expansion behaviour
(opposite sign of $\alpha$ along the two orthogonal directions)
that is a major source of thermo-mechanical stress in high power
applications~\cite{feigelson}. In contrast, LIS expands in the
same direction as temperature increases and exhibits a rather weak
temperature dependence. This favorable situation is related to the
specific nature of the weak Li-S bond already evoked in the
previous subsection.

\subsection{\label{subsec: thermo-optic} Thermo-optic coefficients}

We shall again make use of the labels $X\equiv b, Y\equiv a,
Z\equiv c$ denoting the principal optical axes to describe the
change of the index of refraction versus temperature, for a light
polarized along one of these principal axes and propagating along
another principal direction. For a light wave polarized along
direction $p_j$ and propagating along direction $e_i$, we define
the coefficient $\gamma_{ji}$ related to thermal changes in
optical thickness (path) by
\begin{equation} \label{eq: gammaji}
    \gamma_{ji}=\frac{1}{n_{j}L_{i}}\frac{d(n_{j}L_{i})}{dT}
\end{equation}
where $n_{j}$ and $L_{i}$ are respectively the refractive index
and the length of the sample, with $i,j$ = X, Y or Z and $i\neq
j$. The normalized thermo-optic coefficients
$\beta_{j}=\frac{1}{n_{j}}\frac{dn_{j}}{dT}$ are derived from
(\ref{eq: alphai}) and (\ref{eq: gammaji}) and given by
\begin{equation} \label{eq: betaj}
    \beta_{j}=\gamma_{ji}-\alpha_{i}.
\end{equation}
As evident from Eq.(\ref{eq: betaj}), the knowledge on the values
of the thermal expansion is prerequisite for the interferometric
determination of the thermo-optic coefficients at a given
wavelength. For the orthorhombic symmetry of LIS it can be seen
that each $\beta_{j}$ may be obtained for two directions of light
propagation. This has been performed at four laser wavelengths,
and the resulting linear fits to
\begin{equation}
    \beta_{j}(\lambda,T)=a_{1}(\lambda)+a_{2}(\lambda)T \label{eq: betaj-fit}
\end{equation}
where $T$ is measured in $^\circ$C are summarized in
Table~\ref{tab: table1}. The indicated uncertainties in the
parameters $a_{1}$ and $a_{2}$ represent the deviation from the
mean value determined from the two possible propagation directions
$e_i$ for a given orientation $p_j$ of light polarization. At
$\lambda=1064\,$nm and $T=20^{\circ}\,$C, one has
$dn_{X}/dT=3.725\times 10^{-5}\,$K$^{-1}$, $dn_{Y}/dT=4.545\times
10^{-5}\,$K$^{-1}$, $dn_{Z}/dT=4.467 \times 10^{-5}\,$K$^{-1}$.
Compared with the thermo-optic coefficients of
AGS~\cite{updated-thermooptic}, these values are about 5 times
lower. Dispersion relations based on those wavelength-dependent
data will be derived in Subsection~\ref{subsec:
thermo-optic-dispersion}.

\begin{table}
    \caption{\label{tab: table1} Principal normalized thermo-optic coefficients
    $\beta_{j}=\frac{1}{n_{j}}\frac{dn_{j}}{dT}=a_1+a_2T$ of LIS [Eq.~(\ref{eq: betaj-fit})] at four laser
    wavelengths.\\}
    \begin{ruledtabular}
    \begin{tabular}{cccc}
 $\lambda$ ($\mu$m) &  $\beta_{j}\, (^\circ \text{C}^{-1})$  &  $10^{5}a_{1}\, (^\circ \text{C}^{-1})$  & $10^{8}a_{2}\, (^\circ \text{C}^{-2})$ \\
 \hline \\
                    &  $\beta_{X}$  &  3.41 $\pm$ 0.06 & 3.1 $\pm$ 0.3  \\
 0.4765             &  $\beta_{Y}$  &  4.11 $\pm$ 0.07 & 4.6 $\pm$ 0.9 \\
                    &  $\beta_{Z}$  &  4.08 $\pm$ 0.02 & 4.7 $\pm$ 0.4 \\
 \hline
                    &  $\beta_{X}$  &  2.28 $\pm$ 0.07 & 1.5 $\pm$ 0.4  \\
 0.6328             &  $\beta_{Y}$  &  2.76 $\pm$ 0.05 & 2.1 $\pm$ 0.4 \\
                    &  $\beta_{Z}$  &  2.69 $\pm$ 0.01 & 2.4 $\pm$ 0.1 \\
 \hline
                    &  $\beta_{X}$  &  1.75 $\pm$ 0.07 & 1.0 $\pm$ 0.06  \\
 1.0642             &  $\beta_{Y}$  &  2.10 $\pm$ 0.11 & 1.3 $\pm$ 0.5 \\
                    &  $\beta_{Z}$  &  2.03 $\pm$ 0.08 & 1.3 $\pm$ 0.1 \\
 \hline
                    &  $\beta_{X}$  &  1.50 $\pm$ 0.07 & 0.7 $\pm$ 0.17  \\
 3.392              &  $\beta_{Y}$  &  1.83 $\pm$ 0.14 & 1.1 $\pm$ 0.5 \\
                    &  $\beta_{Z}$  &  1.73 $\pm$ 0.03 & 1.0 $\pm$ 0.02 \\
    \end{tabular}
    \end{ruledtabular}
    \end{table}

\section{\label{sec: vibrational} Vibrational properties}
The vibrational properties have been investigated by means of
Raman spectroscopy and infrared reflectivity in polarized light at
room temperature. The group theory analysis shows that, at the
zone centre for the $\beta$-NaFeO$_{2}$ structure, the 48 normal
phonon modes are distributed among the various irreducible
representations of the C$_{2v}^9$ factor group as follows:
\begin{equation} \label{eq: group}
    \Gamma^{vib}=12\text{A}_{1}+12\text{A}_{2}+12\text{B}_{1}+12\text{B}_{2}.
\end{equation}
The three acoustic phonon modes have A$_{1}$, B$_{1}$ and B$_{2}$
symmetries; the 45 remaining (optical) modes are Raman-active.
A$_{1}$ modes will be active when the experimental configuration
selects a diagonal $aa$, $bb$ or
 $cc$ component of the Raman tensor, while the selection of
 $ba$, $ca$ and $cb$ components will respectively allow the
 observation of the optical phonons of A$_{2}$, B$_{1}$ and B$_{2}$ symmetry.
 A$_{2}$-type phonons being infrared inactive, the IR-reflectivity spectra
 should then contain 11 polar modes of each A$_{1}$, B$_{1}$ and B$_{2}$
 symmetries, with dipole moment respectively parallel to $c$, $a$ and
 $b$ crystallographic axes. In theory, every optical phonon may then be observed
 by means of IR-reflectivity and/or Raman-scattering. Nonetheless,
 depending on various factors such as the more or less polar character of bonds, even an active mode may be unobservable.

\begin{figure}[]
    \includegraphics[width=7cm]{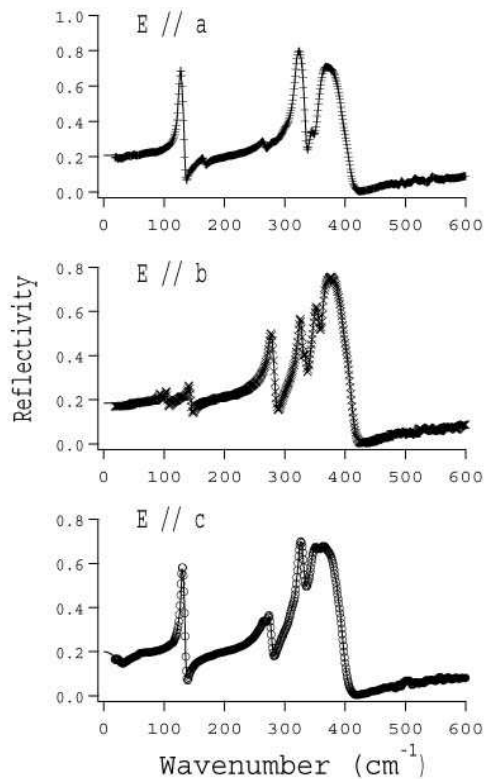}
    \caption{\label{fig: reflect} Infrared reflection curves (symbols)
recorded with polarized light with an electric field parallel to
the crystallographic axes $a$ - B$_{1}$ modes -, $b$ - B$_{2}$
modes - and $c$ - A$_{1}$ modes. The continuous lines represent in
each case the best-fit reflectivity curves calculated with the
model in Eqs.(\ref{eq: permitt})-(\ref{eq: reflectivity}).}
  \end{figure}

To obtain results on every phonon type, we used two LIS rose-tinge
samples cut from the same ingot, with sizes $4\times 4\times
5\,$mm$^{3}$ and parallel optically polished facets (4x4 mm$^{2}$)
normal to $a$ and $c$ axes. Infrared reflection spectra were
recorded in the spectral range 18.3 - 600 cm$^{-1}$, and this
whole range of data was used to adjust the parameters of the
dielectric permittivity model shown in Eq.(\ref{eq: permitt}). The
experimental spectra, recorded at quasi-normal incidence (at an
angle of incidence of $6^{\circ}$) with an electric field
polarization parallel to $a$, $b$ and $c$ axes, are shown in
Fig.~\ref{fig: reflect} (symbols). As can be seen from these
graphs, each reflectivity spectrum is rather complex and exhibits
several more or less overlapping bands, and the high-energy
band(s) in each polarization generally show some additional
structure (slight inflexion in the band shape) that suggests, in
addition to a strong mode, the presence of several other modes
contributing to the same band. Actually, from the numerical
analysis of the experimental results, it is found that for each
direction of polarization, there are three polar modes in the high
frequency band (350 - 420 cm$^{-1}$) and medium frequency band
(250-350 cm$^{-1}$) and the remaining five polar modes, not all
observable in Fig.~\ref{fig: reflect}, are in the low frequency
region ($<250$ cm$^{-1}$).

\begin{figure}[t]
    \includegraphics[width=7cm]{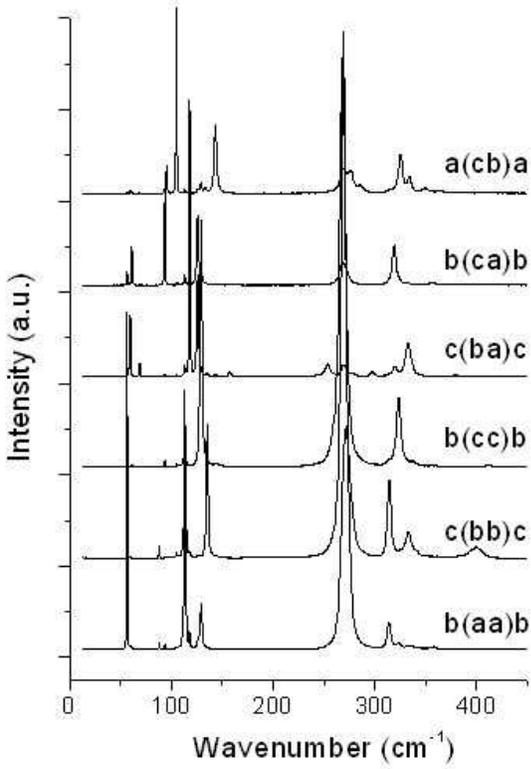}
    \caption{\label{fig: raman} Polarized Raman spectra
(shifted for reasons of clarity) recorded at room temperature. The
experimental configurations $b(aa)b$, $c(bb)c$, $b(cc)b$
correspond to A$_1$ modes, $c(ba)c$ to A$_2$ modes, $b(ca)b$ to
B$_1$ modes and $a(cb)a$ to B$_2$ modes.}
  \end{figure}

Polarized micro-Raman spectra were recorded using a Raman
spectrometer equipped with a confocal microscope and a
liquid-N$_2$ cooled CCD detector in the region 12 - 600 cm$^{-1}$
in triple substractive configuration with an 1800 gr/mm grating.
To increase the quality and resolution of the spectra, we chose to
use the 647nm red line of an argon-krypton laser (P=100 mW) which
experiences lower absorption loss than a green
laser~\cite{photonicwest}). To avoid possible influence of surface
defects, the confocal microscope focused the laser beam at a depth
of 60$\mu$m under the surface. A scattering geometry will be
described by the notation $e_{1}(p_{1}p_{2})e_{2}$~\cite{porto}
where $e_{i}$ identifies the propagation direction of the laser
light, polarized along $p_{i}$, $e$ and $p$ being $a, b$ or $c$
crystallographic axes; the subscripts $i=1$ and 2 refer to
incident and scattered beams respectively. Fig.~\ref{fig: raman}
shows some of the experimental spectra. The Raman spectra are
dominated by a strong mode of A$_{1}$ symmetry at $\sim268$
cm$^{-1}$, also apparent in the IR reflectivity spectra for E//$b$
and E//$c$.

Due to the respective mass of the atoms, high (low) frequency
modes obviously imply movements of Li-S (In-S) bonds. Phonons of
medium frequency (250-350 cm$^{-1}$), visible in both
IR-reflectivity and Raman scattering, are attributed to In-S
bonds~\cite{eifler} while those of highest frequency (above 350
cm$^{-1}$, Fig.~\ref{fig: reflect}) that give rise to strong IR
bands but no observable Raman line are due to weaker Li-S
bonds~\cite{sobota}. Polar modes TO$\approx$LO splitting observed
in the infrared measurements for any of the explored polarizations
is rather important in the high frequency region (strong IR-bands)
which attests for a high polar (ionicity) character of Li-S bonds
in this crystal. The splitting decreases with the phonon frequency
to become so small for low frequency modes that no corresponding
band appears in the reflectivity spectrum which indicates a mostly
covalent character of the In-S bonds. However, even if a phonon
cannot be observed by means of IR-reflectivity, its Raman activity
may be high, which is the case for modes lying below $100$
cm$^{-1}$. Actually intense Raman lines appear in the low and
mid-frequency region, when almost no line can be detected above
roughly 350 cm$^{-1}$. In order to fit the experimental
IR-reflectivity spectra, and get the best-fit reflectivity curves
(lines in Figure~\ref{fig: reflect}, in a good agreement with
experimental points) we used the so-called \emph{four-parameter
model}, convenient to represent even asymmetrical and wide
bands~\cite{salaun}. This model expands the factorized form of the
dielectric permittivity function as
\begin{equation} \label{eq: permitt}
    \varepsilon(\omega)=\epsilon^{'}-i\varepsilon^{"}=
    \varepsilon_{\infty}\prod_{j=1}^{N}\frac{\Omega_{j,\text{LO}}^{2}-\omega^{2}+i\gamma_{j,\text{LO}}\,\omega}
    {\Omega_{j,\text{TO}}^{2}-\omega^{2}+i\gamma_{j,\text{TO}}\,\omega}
\end{equation}
to fit a reflectivity curve described by the Fresnel formula
\begin{equation} \label{eq: reflectivity}
    R(\omega)=\left|\frac{\sqrt{\varepsilon(\omega)}-1}{\sqrt{\varepsilon(\omega)}+1}\right|^{2}.
\end{equation}
In the $\beta$-NaFeO$_{2}$ structure, the model implies for a
given polarization $N$ damped oscillators, the number $N$ being at
most equal to 11 (number of theoretically observable modes,
predicted in this structure by group theory analysis). The $j$-th
damped oscillator represents the $j$-th observed infrared active
couple of modes, characterized by four adjustable parameters,
namely the longitudinal $\Omega_{j,\text{LO}}$ and transverse
$\Omega_{j,\text{TO}}$ angular frequencies, and the corresponding
damping frequencies $\gamma_{j,\text{LO}}$ and
$\gamma_{j,\text{TO}}$. The high-frequency limit values of the
dielectric constant is also adjusted to yield
$\varepsilon_{\infty}$(A$_{1}$)=3.99,
$\varepsilon_{\infty}$(B$_{1}$)=4.09,
$\varepsilon_{\infty}$(B$_{2}$)=3.97.
\begin{table*}[htb]
\caption{\label{tab: table2}Zone-center phonon wavenumbers
($\bar{\nu}_{\text{TO}}/\bar{\nu}_{\text{LO}}$) and damping
($\bar{\gamma}_{\text{TO}}/\bar{\gamma}_{\text{LO}}$), in
cm$^{-1}$, deduced from IR reflectivity and Raman scattering
experiments at room temperature for the various symmetry-adapted
polarization configurations. The transverse (TO) and/or
longitudinal (LO) character of the observed Raman-active phonons
have been determined with the help of the IR reflectivity results
(the notation TO$\approx$LO means that the splitting is too weak
to be unambiguously identified).}
    \begin{center}
    \begin{tabular}{p{1.in}p{1.5in}p{1.5in}p{0.25in}p{1.5in}} \hline
 \multicolumn{1}{c}{Phonon} & \multicolumn{2}{c}{IR reflectivity} &&\multicolumn{1}{c}{Raman}  \\
  \cline{2-3}
 \multicolumn{1}{c}{symmetry} &\multicolumn{1}{c}{Wavenumbers $\bar{\nu}_j$} & \multicolumn{1}{c}{Dampings $\bar{\gamma}_j$} &&
\multicolumn{1}{c}{scattering} \\
 \hline\hline
&\multicolumn{1}{c}{-\,/\,-}&\multicolumn{1}{c}{-\,/\,-}&&\multicolumn{1}{c}{56.4\,(TO$\approx$LO)}\\
&\multicolumn{1}{c}{-\,/\,-}&\multicolumn{1}{c}{-\,/\,-}&&\multicolumn{1}{c}{88.1\,(TO$\approx$LO)}\\
&\multicolumn{1}{c}{-\,/\,-}&\multicolumn{1}{c}{-\,/\,-}&&\multicolumn{1}{c}{112.7/113.3}\\
&\multicolumn{1}{c}{128.6/138.2}&\multicolumn{1}{c}{3.0/4.4}&&\multicolumn{1}{c}{129.1/135.2}\\
\multicolumn{1}{c}{A$_{1}$}
&\multicolumn{1}{c}{266.0/267.1}&\multicolumn{1}{c}{8.5/8.7}&&\multicolumn{1}{c}{268/269}
\\
\multicolumn{1}{c}{TO/LO}
&\multicolumn{1}{c}{274.2/279.2}&\multicolumn{1}{c}{9.2/8.7}&&\multicolumn{1}{c}{273/\,-}
\\
 \multicolumn{1}{c}{(cm$^{-1}$)}
&\multicolumn{1}{c}{274.2/279.2}&\multicolumn{1}{c}{9.2/8.7}&&\multicolumn{1}{c}{273/\,-}
\\
&\multicolumn{1}{c}{313.6/313.8}&\multicolumn{1}{c}{4.9/5.0}&&\multicolumn{1}{c}{314\,(TO$\approx$LO)}
\\
&\multicolumn{1}{c}{323.7/333.7}&\multicolumn{1}{c}{6.2/11.3}&&\multicolumn{1}{c}{323.4/333}
\\
&\multicolumn{1}{c}{341.7/356.0}&\multicolumn{1}{c}{14.3/13.9}&&\multicolumn{1}{c}{341/\,-}
\\
&\multicolumn{1}{c}{357.0/373.0}&\multicolumn{1}{c}{14.5/14.7}&&\multicolumn{1}{c}{357/\,-}
\\
&\multicolumn{1}{c}{373.1/399.0}&\multicolumn{1}{c}{14.3/16.5}&&\multicolumn{1}{c}{-\,/399}
\\ \hline
&&&&\multicolumn{1}{c}{59.4} \\ &&&&\multicolumn{1}{c}{68.7}\\
&&&&\multicolumn{1}{c}{117.8} \\&&&&\multicolumn{1}{c}{125.9} \\
\multicolumn{1}{c}{A$_{2}$}&&&&\multicolumn{1}{c}{157.4} \\
\multicolumn{1}{c}{(cm$^{-1}$)} & \multicolumn{2}{c}{IR inactive}
&&
\multicolumn{1}{c}{253.5} \\ &&&& \multicolumn{1}{c}{297.4} \\
&&&&\multicolumn{1}{c}{319.4} \\ &&&& \multicolumn{1}{c}{332.5} \\
\hline

&\multicolumn{1}{c}{-\,/\,-}&\multicolumn{1}{c}{-\,/\,-}&&\multicolumn{1}{c}{61.1\,(TO$\approx$LO)}\\
&\multicolumn{1}{c}{-\,/\,-}&\multicolumn{1}{c}{-\,/\,-}&&\multicolumn{1}{c}{93.5\,(TO$\approx$LO)}\\
&\multicolumn{1}{c}{125.8/133.4}&\multicolumn{1}{c}{2.6/3.0}&&\multicolumn{1}{c}{125.8/133.3}\\
\multicolumn{1}{c}{B$_{1}$}
&\multicolumn{1}{c}{164.9/166.0}&\multicolumn{1}{c}{8.4/9.0}&&\multicolumn{1}{c}{165\,(TO$\approx$LO)}
\\
\multicolumn{1}{c}{TO/LO}
&\multicolumn{1}{c}{264.9/266.3}&\multicolumn{1}{c}{9.3/10.3}&&\multicolumn{1}{c}{267\,(TO$\approx$LO)}
\\
 \multicolumn{1}{c}{(cm$^{-1}$)}
&\multicolumn{1}{c}{303.3/303.7}&\multicolumn{1}{c}{9.4/9.2}&&\multicolumn{1}{c}{-\,/\,-}
\\
&\multicolumn{1}{c}{318.3/334.6}&\multicolumn{1}{c}{5.0/7.2}&&\multicolumn{1}{c}{318.7/\,-}
\\
&\multicolumn{1}{c}{343.6/347.8}&\multicolumn{1}{c}{11.2/10.4}&&\multicolumn{1}{c}{-\,/\,-}
\\
&\multicolumn{1}{c}{356.4/375.9}&\multicolumn{1}{c}{10.4/22.8}&&\multicolumn{1}{c}{356.5/\,-}
\\
&\multicolumn{1}{c}{376.7/395.8}&\multicolumn{1}{c}{21.9/23.1}&&\multicolumn{1}{c}{-\,/\,-}
\\
&\multicolumn{1}{c}{400.4/408.6}&\multicolumn{1}{c}{29.7/12.4}&&\multicolumn{1}{c}{-\,/\,-}
\\ \hline

&\multicolumn{1}{c}{-\,/\,-}&\multicolumn{1}{c}{-\,/\,-}&&\multicolumn{1}{c}{59.3\,(TO$\approx$LO)}\\
&\multicolumn{1}{c}{96.1/96.4}&\multicolumn{1}{c}{3.3/3.1}&&\multicolumn{1}{c}{96\,(TO$\approx$LO)}\\
&\multicolumn{1}{c}{104.7/105.7}&\multicolumn{1}{c}{4.8/5.2}&&\multicolumn{1}{c}{104.7/105.5}\\
\multicolumn{1}{c}{B$_{2}$}
&\multicolumn{1}{c}{132.0/132.5}&\multicolumn{1}{c}{10.2/10.1}&&\multicolumn{1}{c}{133\,(TO$\approx$LO)}
\\
\multicolumn{1}{c}{TO/LO}
&\multicolumn{1}{c}{142.7/144.9}&\multicolumn{1}{c}{6.1/6.9}&&\multicolumn{1}{c}{143.2\,(TO$\approx$LO)}
\\
 \multicolumn{1}{c}{(cm$^{-1}$)}
&\multicolumn{1}{c}{276.5/285.1}&\multicolumn{1}{c}{8.0/8.4}&&\multicolumn{1}{c}{276.2/285}
\\
&\multicolumn{1}{c}{324.6/329.8}&\multicolumn{1}{c}{5.9/7.3}&&\multicolumn{1}{c}{325.2\,(TO$\approx$LO)}
\\
&\multicolumn{1}{c}{333.4/336.0}&\multicolumn{1}{c}{7.2/7.5}&&\multicolumn{1}{c}{334\,(TO$\approx$LO)}
\\
&\multicolumn{1}{c}{348.0/358.4}&\multicolumn{1}{c}{10.3/8.9}&&\multicolumn{1}{c}{348/\,-}
\\
&\multicolumn{1}{c}{362.2/397.7}&\multicolumn{1}{c}{8.6/11.0}&&\multicolumn{1}{c}{362/\,-}
\\
&\multicolumn{1}{c}{398.6/410.4}&\multicolumn{1}{c}{11.1/11.9}&&\multicolumn{1}{c}{-\,/\,-}
\\ \hline
    \end{tabular}
    \end{center}
\end{table*}

Table~\ref{tab: table2} lists the wavenumbers
$\bar{\nu}_{j}=\Omega_{j}/2\pi c$ and damping rates
$\bar{\gamma}_{j}=\gamma_{j}/2\pi c$ (expressed in cm$^{-1}$) of
the observed optical phonons deduced from IR-reflectivity and/or
Raman scattering experiments, assigned to the various irreducible
representations of the C$_{2v}^9$ factor group. These data allow
to compute the linear dielectric properties of LIS in the far-IR
range, using Eq.(\ref{eq: permitt}). Typically, TO and LO
frequencies deduced from IR-reflectivity experiments are
determined to better than 1 cm$^{-1}$. The precision of a
Raman-active phonon frequency is also generally better than 1
cm$^{-1}$, except for broad and low-intensity lines for which it
decreases to 2 cm$^{-1}$. The transverse or longitudinal character
of the Raman-active modes for A$_1$, B$_1$ and B$_2$ symmetries
has been determined with the help of the IR reflectivity results
where this TO or LO character is immediate. Non-observed IR-active
modes indicate a small TO$\approx$LO splitting and generally did
not allow an accurate determination of TO and LO frequencies from
Raman spectra, but only of a mean value labelled by TO$\approx$LO.
A few Raman lines have been attributed to an interference by an
intense line active in another geometry. These Raman results
complement those from previous studies performed with
unpolarized~\cite{eifler,kovach,bruckner} and polarized
light~\cite{dorday}.

The accurate determination of the frequencies in Table~\ref{tab:
table2} allows to estimate the interatomic bonding force constants
in LIS, using the theoretical considerations developed by Neumann
about the correlation between the vibrational modes of compounds
with sphalerite, chalcopyrite and $\beta$-NaFeO$_{2}$
structures~\cite{neuman-mod}. According to this model the
vibrational spectrum of a crystal with $\beta$-NaFeO$_{2}$
structure contains two sphalerite-like modes the TO-frequencies of
which are nearly independent of the polarization direction.
Further, one of these sphalerite-like modes always corresponds to
the highest $\bar{\nu}_{\text{TO}}$ frequency found in the IR
reflectivity spectra. If the interaction between the atoms is
restricted to next-nearest neighbors which is justified by the
high bond ionicities in Li -containing ternary compounds, the
wavenumbers of the transverse optical modes of these
sphalerite-like modes in LiB$^{\text{III}}$C$_{2}^{\text{VI}}$ are
given by~\cite{neuman-nu}
\begin{equation} \label{eq: nu-TO}
    \bar{\nu}_{TO,k}^{2}=\frac{1}{\pi^{2}c^{2}}(\alpha_{k}-\delta\alpha_{k})
    \left(\frac{1}{m_{k}}+\frac{1}{m_{\text{C}}}\right)
\end{equation}
with the superscript $k=$ in the first parenthesis denoting Li-C
or B-C cation-anion bonds (in the present case, $C=$S and $B=$In).
Here $c$ is the speed of light, $\alpha_{k}$ the mechanical
bond-stretching force constant, $m_{k}$ the mass of the cation and
$m_{\text{C}}$ the mass on the chalcogen anion (C=S$^{2-}$). The
quantity $\delta\alpha_{k}$ accounts for the fact that in the case
of IR active modes the local atomic displacements give rise to a
non-vanishing net dipole moment per unit cell and thus, to
additional dipole-dipole interactions. It is given by
$\delta\alpha_{k}=\sqrt{3}(e_{k})^{2}/(16\varepsilon_{0}s_{k}^{3})$,
where $s_{k}$ is the average bond length ($s_{Li-S}=0.2438$ nm,
$s_{In-S}=0.2452$ nm), $e_{k}=0.28ef_{k}(Z_{k}+Z_{C})$ is an
effective charge depending on the bond ionicity $f_{k}$ and the
valences of the cation and anion participating to the bond. In
Ref.~\cite{sobota}, from an unpolarized IR reflectivity spectrum
of polycrystalline LIS restricted to the region 200 - 400
cm$^{-1}$, the highest-frequency TO/LO mode identified corresponds
to 356/406 cm$^{-1}$ and the second TO/LO sphalerite-like mode to
323/334 cm$^{-1}$. Actually from the inspection of Table~\ref{tab:
table2}, the highest-frequency sphalerite-like modes with nearly
identical frequencies in B$_{1}$ and B$_{2}$ symmetries are
respectively the 400.4/408.6 cm$^{-1}$ and 398.6/410.4 cm$^{-1}$
(this mode is not observed in A$_{1}$). The second sphalerite-like
mode is recognized from Table~\ref{tab: table2} as the 323.7/333.7
cm$^{-1}$ (A$_{1}$), 318.3/334.6 cm$^{-1}$ (B$_{1}$), 324.6/329.8
cm$^{-1}$ (B$_{2}$), close to the value identified in
Ref.~\cite{sobota}. The Li-S bond force constant deduced from
Eq.(\ref{eq: nu-TO}) by Sobotta \emph{et al}~\cite{sobota},
$\alpha_{Li}=24.7$ N/m, is then underestimated by
$(400.4/356)^{2}=24$\% while the In-S bond force constant, deduced
from the second sphalerite-like mode remains unchanged to
$\alpha_{In}=53.4$ N/m. But even with this slight correction, the
Li-S bond remains still about twice weaker than the In-S bond. In
contrast the strength of the Ag-C$^{\text{VI}}$ bond in
chalcopyrite compounds is more comparable to that of the
B$^{\text{III}}$-C$^{\text{VI}}$ bonds~\cite{sobota}, underlining
the specific nature of the lithium-chalcogen bond in crystals with
tetrahedral coordination of the atoms. As a consequence, the
increased lattice phonon energy of LIS, as compared with AGS or
AGSe, favors heat dissipation, hence a higher thermal conductivity
is expected. Indeed, an unpublished measurement of LIS thermal
conductivity yielded $K_{a}=6.0$ W/(m$\cdot$K), $K_{b}=6.2$
W/(m$\cdot$K) and $K_{c}=7.6$ W/(m$\cdot$K)~\cite{ebbers}, i.e. 5
times larger than for AGS and of the same magnitude as for
LiNbO$_{3}$.

A last important feature from the vibrational spectra is that the
phonon spectrum for LIS is located at wavenumbers below 410
cm$^{-1}$. The IR wavelength cut-off (8.1 $\mu$m) in
Fig.~\ref{fig: ftir} quite exactly corresponds to three times this
highest energy phonon. The long wavelength shape of the
transmission curves - and limit of the transparency range - seems
then to be simply due to multiphonon processes. It is worth
mentioning that, although the vibrational investigations have used
rose tinge samples, the phonon spectrum should not depend on
coloration since all LIS samples display the same IR cutoff edge,
with or without annealing.

\section{\label{sec: electro-piezo} Piezo-electric and electro-optic coefficients}

\subsection{\label{subsec: apparent-eo} Apparent electro-optic coefficients}
The measurement of electro-optic coefficients involves the
determination of changes in optical thickness when applying an
electric field to the material. LIS belongs to the symmetry class
$mm2$ for which the electro-optic tensor exhibits 5 nonzero
components, namely $r_{13}$, $r_{23}$, $r_{33}$, $r_{42}$ and
$r_{51}$~\cite{ire-standard}. In this subsection we will describe
the electro-optic effect in the principal optic frame where $XYZ$
are the axes of the index ellipsoid which as we shall see in the
next section coincide with the crystallographic $bac$ axes. We
will address only the $r_{i3}$'s coefficients that can be obtained
by applying a field $E_{3}$ along the $Z$-axis which does not lead
to a rotation of the ellipsoid. Optical interferometric methods
may be advantageously used in this case for their determination.
An applied electric field $\mathbf{E}$, with respective components
$(E_{1}, E_{2}, E_{3})$ along the principal axes $(X, Y, Z)$, will
modify the index ellipsoid according to
\begin{eqnarray}
    (n_{X}^{-2}+r_{13}E_{3})X^{2}+(n_{Y}^{-2}+r_{23}E_{3})Y^{2} \nonumber\\
    +(n_{Z}^{-2}+r_{33}E_{3})Z^{2}
    +2r_{42}E_{2}YZ+2r_{51}E_{1}XZ=1.
\end{eqnarray}
The new indices $n_{i}(E_{3})$ ($i\equiv1,2,3\equiv X,Y,Z$) are
given by
\begin{equation}
    n_{i}(E_{3})=n_{i}-\frac{1}{2}n_{i}^{3}r_{i3}E_{3}
\end{equation}
where $\delta(n_{i})=-\frac{1}{2}n_{i}^{3}r_{i3}E_{3}$ represents
the variation of the refractive index $n_{i}$.

Due to the piezo-electric effect the field $E_{3}$ will also
generate a variation $\delta(L_{j})$ of the crystal thickness
$L_{j}$ along a given direction $e_j$ such that
\begin{equation} \label{eq: dej}
    \delta(L_{j})=L_{j}\bar{d}_{3j}E_{3}
\end{equation}
where $\bar{d}_{3j}$ ($j=1,2,3$) are components of the
piezo-electric tensor~\cite{ire-standard}. For a light wave
propagating along direction $e_j$ and polarized along direction
$p_i$ of the crystal the variation $\delta(n_{i}L_{j})$ of the
optical path $n_{i}L_{j}$ induced by an applied electric field
$E_{3}$ will be
\begin{equation}
    \delta(n_{i}L_{j})=n_{i}L_{j}\left(\bar{d}_{3j}-\frac{1}{2}n_{i}^{2}r_{i3}\right)E_{3}.
\end{equation}
We define the \emph{effective} electro-optic coefficients as
\begin{equation} \label{eq: ri3-eff}
    r_{i3}^{\text{eff}}=-\frac{2}{n_{i}^{2}}\left(\bar{d}_{3j}-\frac{1}{2}n_{i}^{2}r_{i3}\right)
\end{equation}
with $j=1,2$, which represent in fact the (apparent) electro-optic
coefficients obtained in most experiments where the piezo-electric
contribution is ignored. The measurements of the
$r_{i3}^{\text{eff}}$'s were performed by using the technique of
Phase-modulated Thermal Scanning Interferometry
(PTSI)~\cite{sandrine-these}. In this method, as for the
measurements of thermo-optic coefficients, the sample acts as a
Fabry-Perot thermal scanning interferometer in which phase
modulation is achieved by applying an AC field $E_{3}$ along the Z
axis. Various configurations of light polarization $(E_{X}, E_{Y},
E_{Z})$ and wave vector $\mathbf{k}$ $(k_{X}, k_{Y})$ were used to
determine the apparent electro-optic coefficients of LIS.

The two LIS samples of aperture $5\times 5\,$mm$^2$ used for the
thermo-optic data measurements with lengths $L=8\,$mm along the X
and Y propagation axes were supplied with gold evaporated
electrodes on their Z facets. The low frequency (480 Hz) applied
voltage was 200 V peak-to-peak and linear temperature ramps of
$0.2^{\circ}$C/min were used from $20^{\circ}$C to $55^{\circ}$C.
Measurements of apparent electro-optic coefficients were performed
at $1.064\,\mu$m by using a 100mW TEM$_{00}$ single-frequency
Nd:YAG laser. In the scanned temperature interval a constant value
has been observed for each of them; they are summarized in
Table~\ref{tab: table3}.
\begin{table}
\caption{\label{tab: table3} Apparent electro-optic coefficients
of LIS; 3(X) and 3(Y) represent the values of
$r_{33}^{\text{eff}}$ obtained for a light propagation along X and
along Y, respectively.\\}
\begin{ruledtabular}
\begin{tabular}{cc}
 $i$ &  $r_{i3}^{eff}$ (pm/V) \\
 \hline \\
       1            &  2.18$\pm$0.04   \\
       2            &  2.82$\pm$0.33   \\
       3(X)         &  0.93$\pm$0.03  \\
       3(Y)         &  0.11$\pm$0.02  \\
\end{tabular}
\end{ruledtabular}
\end{table}
These data feature a quite weak electro-optic effect in LIS and
the large discrepancy observed between $r_{i3(X)}^{\text{eff}}$
and $r_{i3(Y)}^{\text{eff}}$ suggests a strong contribution of the
piezo-electric effect to the apparent electro-optic one.

\subsection{\label{subsec: pure-eo} Piezo-electric and pure electro-optic coefficients}

The non-vanishing piezo-electric coefficients $\bar{d}_{31}$,
$\bar{d}_{32}$ and $\bar{d}_{33}$ were measured by using the
modified Mach-Zehnder arrangement of the absolute interferometric
dilatometer in a manner similar to the one employed for dilatation
measurements. Phase modulation of
the interference pattern is performed by applying to the samples
the same voltage and same thermal scanning procedure as described
previously. The algebraic sign of the coefficients was attributed
according to the ANSI/IEEE Standard 176-1987 on piezo-electric
crystals~\cite{ire-standard} and we have found in pm/V units and
in the XYZ frame: $\bar{d}_{31}=-5.6\pm 0.1$ ,
$\bar{d}_{32}=-2.7\pm 0.2$ and $\bar{d}_{33}=+7.5\pm 0.2$.

From Equation (\ref{eq: ri3-eff}) we determine the pure
electro-optic coefficients $r_{13}=1.0\,$pm/V, $r_{23}=0.4\,$pm/V
and $r_{33}=-1.3\,$pm/V which reveals a rather small true
electro-optic effect compared to the piezo-electric one. The
relative uncertainty in these values is close to 15\%. These
measurements confirm that, as AGS or AGSe, LIS is a poor candidate
for electro-optic or piezo-electric based devices. A comparison of
our results with earlier data of Negran \emph{et al}~\cite{negran}
cannot be meaningful since the $r_{13}$, $r_{23}$ and $r_{33}$
that were measured there did not account for the piezo-electric
contribution and were not determined independently, but as a
linear combination.

\subsection{\label{subsec: soref} Comparison of the electro-optic and pyroelectric coefficients}

From the anharmonic oscillator model developed by
Garrett~\cite{garrett}, Soref~\cite{soref} yields a semi-empirical
relationship between the pyroelectric coefficient at constant
stress ($p^{\,\sigma}=\left(\frac{\Delta P_{s}}{\Delta
T}\right)_{\sigma}$) and the second-order electronic
susceptibility $\chi^{(2)}$ describing the linear (true)
electro-optic effect:
\begin{equation}
    p^{\,\sigma}=\chi^{(2)}(\omega\pm0;\omega,0)2C_{v}\frac{\epsilon_{0}m_{e}^{2}(\omega_{e}^{2}-\omega^{2})^{2}}{\mu\omega_{i}^{2}N_Ae^{2}}\left(\frac{3A+B}{C+3D}\right).
\end{equation}
The pyroelectric coefficient $p^{\sigma}$ gives the variation of
the spontaneous polarization $\Delta P_{s}$  induced by a
temperature variation $\Delta T$. The parameters $m_{e}$ and $e$
are the mass and charge of the electron, $\omega_{e}$ and
$\omega_{i}$ are respectively the electronic and ionic resonance
frequencies, $C_{v}$ is the molar specific heat at constant volume
(which differs from the specific heat at constant pressure $C_{p}$
measured in Subsection~\ref{subsec: specheat} by a negligible
quantity) and $N_A$ the Avogadro number. The constant terms $A, B,
C$ and $D$ are the coefficients related to the third order
expansion of the crystal potential energy in the electronic and
ionic coordinate system $q_{e}$ and $q_{i}$~\cite{soref}. The
ratio $(3A+B)/(C+3D)$ indicates the extent to which the
predominant ionic contribution to the anharmonic potential at
$\omega\approx \omega_{i}$ differs from the predominant electronic
one at $\omega>>\omega_{i}$. This ratio is expected to be constant
whatever the material, suggesting a linear relationship between
the electro-optic and pyroelectric coefficients. This hypothesis
was verified for a few known materials, mostly of the oxide
type~\cite{soref}.

By using $C_{v}=C_{p}$ and $\chi_{333}^{(2)}=n_{3}^{4}r_{33}$ from
the previous subsection, the pyroelectric coefficient measured
previously~\cite{bidault} as a function of temperature and its
value extrapolated at 300 K, $p^{\,\sigma}=6\times10^{-10}$
C$\cdot$ cm$^{-2}\,$K$^{-1}$, we observe that Soref's hypothesis
is confirmed for the ternary semiconductor LIS, as it was already
shown for the binary one CdS~\cite{soref}. We note also that a
rather good estimate of the pyroelectric effect may be deduced
from electro-optic measurements and vice versa, but this requires
at first an accurate determination of the piezo-electric
contribution.

\section{\label{sec: linear} Linear optical and thermo-optic dispersions}
We now address, in light of the vibrational spectra results, the
most important data for a nonlinear material: linear optical and
thermo-optic dispersion relations that allow to predict accurately
the phase-matching directions and temperature tunability.
\subsection{\label{subsec: sellmeier}  Sellmeier equations at room-temperature}
The only available dispersion data of LIS, measured using the
minimum deviation technique, were compiled by Boyd \emph{et al} in
1973~\cite{boyd} and fitted to one-pole (UV) Sellmeier
equations~\cite{ebbers-report} reproduced also in
Refs.~\cite{rotermund,grechin,grechin2}. Although the accuracy of
their index determination was not reported, from subsequent
nonlinear phase-matched conversion experiments it appeared that
the Sellmeier equations based on their data are accurate to
predict the birefringence (via angle acceptance bandwith
measurements~\cite{assl}), but fail to predict accurately the
absolute phase-matching directions. Discrepancies of several
degrees from the expected phase-matching angles were
reported~\cite{knippels}. Such deviations cannot be attributed to
the X-ray sample orientation uncertainties ($\pm0.5^{\circ}$).
Furthermore, we have checked that the phase-matching angles, and
hence the linear refractive indices, are independent of crystal
coloration or exact stoichiometry.

Following the standard convention for assigning the principal axes
($X, Y, Z$) to the crystallographic axes ($a, b, c$), e.g.
$n_{X}<n_{Y}<n_{Z}$, one has $X\leftrightarrow b$,
$Y\leftrightarrow a$ and $Z\leftrightarrow c$ for LIS since
$n_{b}<n_{a}<n_{c}$. A phase-matching direction will be labelled
by ($\theta, \varphi$), where $\theta$ is the angle of the
wavevector respective to the polar $Z$ axis and $\varphi $ is the
azimuthal angle between its projection in the $X-Y$ plane and the
$X$ axis.  To refit the Sellmeier equations given in
Ref.~\cite{ebbers-report}, we have used the experimental type-II
(eoe, oeo) SHG angles measured in the fundamental wavelength range
$\lambda\in [2.5 - 6]\,\mu$m~\cite{knippels,photonicwest},
supplemented with new recent data derived from type-II (eoe)
difference-frequency generation (DFG) in the X-Y plane (see
subsection~\ref{subsec: dfg}), where in the wavelength-tuning
mode, mid-IR down-conversion in the range $\lambda_1\in
[6.6-7]\,\mu$m was achieved at normal incidence. Table~\ref{tab:
table4} compiles the experimental phase-matching data used to
derive the following room-temperature ($T=20^\circ$C) two-pole
Sellmeier equations, valid over the full transparency window,
using a nonlinear least-square fit of the data,
\begin{eqnarray}
n_{X}^{2}=6.686059 +
\frac{0.1385833}{\lambda^{2}-0.05910334}+\frac{2047.46509}{\lambda^{2}-897.7476},
\label{eq: sellmz1}
\\
n_{Y}^{2}=7.095493 +
\frac{0.1422326}{\lambda^{2}-0.06614640}+\frac{2511.08936}{\lambda^{2}-988.2024},
\label{eq: sellmz2}
\\
n_{Z}^{2}=7.256327 +
\frac{0.15072}{\lambda^{2}-0.06823652}+\frac{2626.10840}{\lambda^{2}-983.0503}.
\label{eq: sellmz3}
\end{eqnarray}

\begin{table*}[ht]
\caption{\label{tab: table4} Phase-matching conditions for type-II
SHG and DFG in LiInS$_{2}$.\\\\}
\begin{ruledtabular}
\begin{tabular}{cccccc}
 &  \multicolumn{3}{c}{Wavelength (nm)\footnote{$1/\lambda_{1}+1/\lambda_{2}=1/\lambda_{3}$ with $\lambda_1\geq \lambda_2>\lambda_3$.}} & \multicolumn{2}{c}{Phase-Matching Angles ($\theta,\varphi$)\footnote{In degree.}} \\
\cline{2-4} \cline{5-6}
 & $\lambda_{1}$ & $\lambda_{2}$ & $\lambda_{3}$ & Measured & Calculated \\
\hline
 SHG      & 2366   & 2366   & 1183   &  (90,\,82.1) & (90,\,83.908) \\
(XY: $eoe$)& 2469   & 2469   & 1234.5 & (90,\,73.1) & (90,\,72.755) \\
          & 2527   & 2527   & 1263.5 &  (90,\,69.8) & (90,\,69.235) \\
          & 2583   & 2583   & 1291.5 &  (90,\,67.4) & (90,\,66.507) \\
          & 2900   & 2900   & 1450   &  (90,\,57.9) & (90,\,56.884) \\
          & 3400   & 3400   & 1700   &  (90,\,50.7) & (90,\,50.543) \\
          & 3700   & 3700   & 1850   &  (90,\,48.3) & (90,\,49.392) \\
          & 3900   & 3900   & 1950   &  (90,\,49.0) & (90,\,49.381) \\
          & 4450   & 4450   & 2225   &  (90,\,51.6) & (90,\,51.772) \\
          & 4950   & 4950   & 2475   &  (90,\,56.3) & (90,\,56.543) \\
          & 5350   & 5350   & 2675   &  (90,\,62.8) & (90,\,62.227) \\
          & 5550   & 5550   & 2775   &  (90,\,66.0) & (90,\,65.916) \\
          & 5750   & 5750   & 2875   &  (90,\,69.1) & (90,\,70.508) \\
          & 5900   & 5900   & 2950   &  (90,\,71.4) & (90,\,75.006) \\
\hline \\
 SHG      & 2542.7 & 2542.7 & 1371.35 & (35.4,\,90) & (36.452,\,90)\\
(YZ: $eoo$)& 2552.7 & 2552.7 & 1276.35 & (34.0,\,90)& (34.684,\,90)\\
          & 2570.4 & 2570.4 & 1285.2  & (31.0,\,90) & (31.508,\,90)\\
          & 2587   & 2587   & 1293.5  & (28.7,\,90) & (28.435,\,90)\\
          & 2602.3 & 2602.3 & 1301.15 & (25.9,\,90) & (25.481,\,90)\\
          & 2606.7 & 2606.7 & 1303.35 & (25.1,\,90) & (24.592,\,90)\\
          & 2631.4 & 2631.4 & 1315.7  & (19.6,\,90) & (19.211,\,90)\\
\hline \\
 DFG      & 6585.14 & 872.240 & 770.220 & (90,\,42) & (90,\,42.170)\\
(XY: eoe) & 6623.95 & 865.801 & 765.716 & (90,\,42) & (90,\,42.145)\\
          & 6719.87 & 850.443 & 754.905 & (90,\,42) & (90,\,42.076)\\
          & 6762.44 & 843.761 & 750.162 & (90,\,42) & (90,\,42.046)\\
          & 6832.63 & 833.187 & 742.629 & (90,\,42) & (90,\,41.992)\\
          & 6899.92 & 822.391 & 734.810 & (90,\,42) & (90,\,41.967)\\
          & 6989.56 & 810.827 & 726.544 & (90,\,42) & (90,\,41.863)\\
\end{tabular}
\end{ruledtabular}
\end{table*}

  \begin{figure}[b]
    \includegraphics[width=7cm]{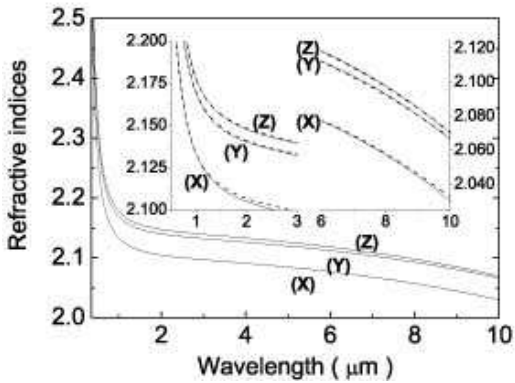}
    \caption{\label{fig: indices}  Principal refractive
    indices of LiInS$_{2}$ as computed from Eqs.(\ref{eq:
    sellmz1})-(\ref{eq: sellmz3}). The inset plots show the UV and
    mid-IR portions on an expanded scale and also a comparison with a
    calculation based on the Sellmeier expansions from
    Ref.~\cite{ebbers-report} (dashed lines).}
  \end{figure}

In Eqs.(\ref{eq: sellmz1})-(\ref{eq: sellmz3}), the wavelengths
are expressed in microns. The three IR wavelength poles
$\lambda_{IR}(X)=29.96\,\mu$m, $\lambda_{IR}(Y)=31.44\,\mu$m,
$\lambda_{IR}(Z)=31.35\,\mu$m (i.e.
$\bar{\nu}(X)=333.7\,$cm$^{-1}$, $\bar{\nu}(Y)=318.1\,$cm$^{-1}$,
$\bar{\nu}(Z)=318.9\,$cm$^{-1}$) correspond to the
medium-frequency phonon band-centers of symmetry B$_{2}$, B$_{1}$
and A$_{1}$ respectively, attributed to the In-S vibrations (see
Table~\ref{tab: table2}). We note that another two-pole Sellmeier
set of equations that recently appeared~\cite{badikov2} is based
on the old index data of Boyd et al.~\cite{boyd} while the mid-IR
poles were not associated with phonon frequencies. Fig.~\ref{fig:
indices} displays the principal indices of refraction deduced from
Eqs.(\ref{eq: sellmz1})-(\ref{eq: sellmz3}). No index crossing
($n_{o}(\lambda)=n_{e}(\lambda)$) is found in the blue part of the
dispersion data, in contrast with the uniaxial
AGS~\cite{updated-thermooptic}. The two optic axes, lying in the
X-Z plane, determine the propagation directions where the index of
refraction is independent of the polarization. The angle between
the two optic axes is $2V_{Z}$ where $V_{Z}$ is the angle of each
axis with the polar Z-axis. $V_{Z}$ is wavelength-dependent and
can be calculated from~\cite{dmitriev}
\begin{equation} \label{eq: optic-axis}
    \sin
    V_{Z}=\frac{n_{Z}(n_{Y}^{2}-n_{X}^{2})^{1/2}}{n_{Y}(n_{Z}^{2}-n_{X}^{2})^{1/2}}.
\end{equation}

\begin{figure}[tb]
    \includegraphics[width=7cm]{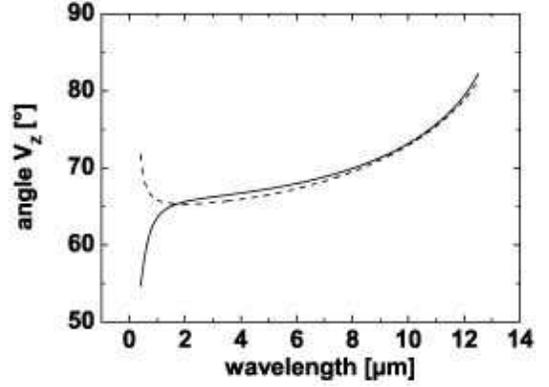}
    \caption{\label{fig: optic-axis}  Angle $V_{Z}$ between the optic axes and the Z-axis in LIS calculated with the Sellmeier
     expansion of Ref.~\cite{ebbers-report} (dashed line), and the new Sellmeier expansion (Eqs.(\ref{eq: sellmz1})-(\ref{eq: sellmz3})) (solid line).}
\end{figure}

Fig.~\ref{fig: optic-axis} plots $V_{Z}$ as a function of the
wavelength, using Eqs.(\ref{eq: sellmz1})-(\ref{eq: sellmz3})
(solid line) and using the Sellmeier equations of
Ref.~\cite{ebbers-report} (dashed line). Let us note the different
behavior for $\lambda < 1\,\mu$m (absence of a retracing
behavior). Note that although the principal indices on which the
old dispersion relations are based were measured from the minimum
deviation technique in 1973~\cite{boyd}, the absorption
coefficient of the prism samples used then was quite different
(much larger than in Fig.~\ref{fig: ftir} up to about $2.5\,\mu$m)
which obviously is the reason for the difference observed in
Fig.~\ref{fig: optic-axis}. Since $V_{Z}>45^{\circ}$ LIS is a
negative biaxial crystal. At longer wavelengths it can be regarded
as quasi-uniaxial, which is a consequence of the fact that the two
indices $n_{Y}$ and $n_{Z}$ are very close. Therefore no
phase-matching can be expected for propagation directions in the
vicinity of the X-principal axis.

\begin{figure}[tb]
    \includegraphics[width=7cm]{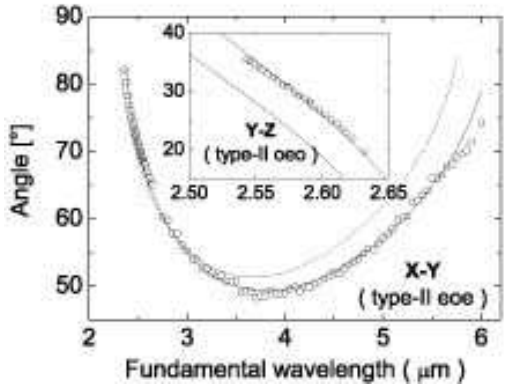}
    \caption{\label{fig: shg-pm}  Phase-matching angles for type-II ($eoe$ or $oee$) SHG in the X-Y principal plane (main frame)
    and type-II ($oeo$ or $eoo$) SHG in the Y-Z plane (inset frame) of LIS. The fundamental source for
    $\lambda<2.65\,\mu$m ($\lambda>2.75\,\mu$m, respectively)
    is the idler wave of a Nd:YAG-pumped nanosecond LiNbO$_{3}$ parametric oscillator (the frequency-doubled radiation from a Free-Electron Laser). The symbols refer to experiments while the solid lines are calculated from
    Eqs.(\ref{eq: sellmz1})-(\ref{eq: sellmz3}). The dashed lines are computed from the Sellmeier equations of Refs.~\cite{ebbers-report}.}
  \end{figure}

The agreement between the experimental phase-matching data used to
derive Eqs.(\ref{eq: sellmz1})-(\ref{eq: sellmz3}) and the
calculated ones is illustrated in Fig.\ref{fig: shg-pm} (solid
lines) and Table~\ref{tab: table4}. The accuracy of phase-matching
angle prediction is $\pm 1.5^{\circ}$. The eight last SHG points
in Fig.~\ref{fig: shg-pm} were discarded from the nonlinear fit
for obvious lack of reliability. Only a two-pole form leads to the
smallest residual discrepancies between calculated and
experimental data, while the former one-pole Sellmeier
equations~\cite{ebbers-report} are unable to predict SHG near the
cut-off wavelength of $\sim 6\,\mu$m.
 Let us note that there is a discrepancy
of $\sim +6^\circ$ between the experimental phase-matching angles
for SHG in the Y-Z plane and the ones calculated with the old
Sellmeier equation, which means that the Y-Z birefringence given
by Eqs. (\ref{eq: sellmz1})-(\ref{eq: sellmz3}) is more accurate.
The previous one-pole Sellmeier equations also predict DFG
wavelengths $\sim +40\,$nm longer than experimentally observed.
Under the usual designation $\lambda_1\geq \lambda_2>\lambda_3$
the FWHM DFG
angular acceptance (FWHM) measured at fixed wavelengths was
$\Delta\varphi=0.19^\circ$ for a crystal length of 10 mm, i.e.
much smaller than the uncertainty on crystal orientation ($\pm
0.5^{\circ}$). Hence the uncertainty ($\pm1\,$nm at maximum) of
wavelength measurement in Table~\ref{tab: table4} cannot explain
the $\sim 40\,$nm shift observed with the previous Sellmeier
equations, confirming the increased accuracy in dispersion
provided by Eqs.(\ref{eq: sellmz1})-(\ref{eq: sellmz3}).

\subsection{\label{subsec: thermo-optic-dispersion} Thermo-optic dispersion relations}

The thermo-optic coefficients $\beta_{j}$ [Eq.~(\ref{eq:
betaj-fit})] are functions of both temperature $T$ (in $^\circ$C)
and wavelength $\lambda$ (in $\mu$m). In analogy with the index
dispersion equations at room-temperature (see previous subsection)
the parameters $a_{k}(\lambda)$ ($k=1, 2$) in Table~\ref{tab:
table1} are fitted according to the following two-pole functional
form,
\begin{equation} \label{eq: beta-coeff}
    a_{k}(\lambda)=C_{0,k}+\frac{C_{1,k}}{\lambda^{2}-\lambda_{01,k}^{2}}+\frac{C_{2,k}}{\lambda^{2}-\lambda_{02,k}^{2}}.
\end{equation}
Table~\ref{tab: table5} compiles the fitting parameters $C_{0,k},
C_{1,k}, C_{2,k}$, $\lambda_{01,k}$ and $\lambda_{02,k}$.
\begin{table*}[t]
\caption{\label{tab: table5}  Fitting parameters describing the
thermal and wavelength dependence of the principal thermo-optic
coefficients of LIS, according to Eqs.(\ref{eq:
beta-lam})-(\ref{eq: beta-coeff}).\\}
\begin{ruledtabular}
\begin{tabular}{ccccccc}
    $a_{k}(\lambda)$ & $p_i$ &         $C_0$         &          $C_1$         &        $C_2$           & $\lambda_{01}$ &$\lambda_{02}$\\
\hline \\
                     & X   & 2.899680$\times10^{-5}$ &0.2033761$\times10^{-5}$&297.0535$\times10^{-5}$ &0.3475601&14.86755\\
       $a_{1}$       & Y   & 2.478267$\times10^{-5}$ &0.2612463$\times10^{-5}$&122.3207$\times10^{-5}$ &0.3391212&13.91854\\
                     & Z   & 3.315830$\times10^{-5}$ &0.2555208$\times10^{-5}$&364.9813$\times10^{-5}$ &0.3451539&15.44113\\
\hline \\
                     & X   & 1.987083$\times10^{-8}$ &0.1860292$\times10^{-8}$&250.0777$\times10^{-8}$ &0.3886392&14.25991\\
       $a_{2}$       & Y   & 0.1523235$\times10^{-8}$&0.2601335$\times10^{-8}$&-191.8712$\times10^{-8}$&0.3969192&14.79803\\
                     & Z   &-06564806$\times10^{-8}$ &0.4291068$\times10^{-8}$&-343.1015$\times10^{-8}$&0.3435849&14.94811
\end{tabular}
\end{ruledtabular}
\end{table*}
According to the definition
$\beta_{j}=\frac{1}{n_{j}}\frac{dn_{j}}{dT}$, the index of
refraction at a given temperature $T$ will be given by
\begin{eqnarray}
    \frac{n(\lambda,T)}{n(\lambda,T_{0})}&=& \exp\left[a_{1}(\lambda)(T-T_{0})+a_{2}(\lambda)\frac{(T-T_{0})^{2}}{2}\right]
    \nonumber\\
                &\simeq &
                1+a_{1}(\lambda)(T-T_{0})+a_{2}(\lambda)\frac{(T-T_{0})^{2}}{2} \label{eq: nT}
\end{eqnarray}
where $T_{0}=20^{\circ}$C and $n(\lambda,T_{0})$ is given by
Eqs.(\ref{eq: sellmz1})-(\ref{eq: sellmz3}).

These thermo-optic dispersion relations will have to be refined by
experimental temperature phase-matching data in the future.
Relations (\ref{eq: beta-coeff}) with the constant values in
Table~\ref{tab: table5} would provide a suitable basis for such a
refit procedure. They predict a FWHM temperature acceptance of
$\Delta TL=248^{\circ}$C$\cdot$cm ($L$: crystal length) at room
temperature for the type-II (eoe) SHG at the fundamental
$\lambda=2.5\,\mu$m in the X-Y plane. Hence LIS is expected to be
very stable against self-induced thermal effects. As a
counterpart, its temperature tunability is expected to be somewhat
limited. However for the DFG of near-IR sources
($\lambda_{3}=754.905\,$nm, $\lambda_{2}=850.443\,$nm, see
Table~\ref{tab: table4}), the FWHM temperature acceptance is
predicted to be only $\Delta TL=46^{\circ}$C$\cdot$cm.

\section{\label{sec: phase-matching} Phase-matching investigations }

\subsection{\label{subsec: deff} Effective nonlinearity and Hobden classification}
The phase-matching loci and the effective nonlinearity for LIS
will be analyzed in the XYZ-principal optic axis frame, but the
tensor elements of the nonlinear susceptibility are defined
traditionally, in accordance with ANSI/IEEE Std
176-1987~\cite{ire-standard}, in the $abc$-crystallographic frame
where $c$ is the polar two-fold axis as in Ref.~\cite{boyd}. The
chosen convention $n_{X}<n_{Y}<n_{Z}$ in order to have the two
optic axes in the X-Z principal plane affects therefore the
expressions for the effective second order nonlinearity
$d_{\text{eff}}$.

The general form of the contracted $\mathbf{d}^{(2)}$ tensor for
the orthorhombic class $mm2$ reads as

\begin{equation} \label{eq: d-tensor}
\mathbf{d}^{(2)}=\left(
\begin{array}{cccccc}
  0 & 0 & 0 & 0 & d_{15} & 0 \\
  0 & 0 & 0 & d_{24} & 0 & 0 \\
  d_{31} & d_{32} & d_{33} & 0 & 0 & 0 \\
\end{array}%
\right).
\end{equation}

For an arbitrary propagation direction assuming collinear
interaction and neglecting the spatial walk-off effect analytical
expressions for $d_{\text{eff}}$ can be derived in our conventions
from the general formulae presented by Dmitriev \emph{et
al}~\cite{dmitriev2},
\begin{equation}\label{eq: deff-ssf}
    \begin{split}
    d_{\text{eff}}^{\,ssf} &= \;2\,d_{15}\sin\theta \cos\delta\, (\cos\theta \sin\varphi \sin\delta -\cos\varphi \cos\delta ) \\
                           &\times (\cos\theta \sin\varphi \cos\delta +\cos\varphi \sin\delta )\\
                           &+ \;2\,d_{24}\sin\theta \cos\delta\, (\cos\theta \cos\varphi \cos\delta -\sin\varphi \sin\delta ) \\
                           &\times (\cos\theta \cos\varphi \sin\delta +\sin\varphi \cos\delta )\\
                           &+ \;\;\;d_{31}\sin\theta \sin\delta\, (\cos\theta \sin\varphi \cos\delta+\cos\varphi\sin\delta)^{2}\\
                           &+ \;\;\;d_{32}\sin\theta \sin\delta\, (\cos\theta \cos\varphi \cos\delta -\sin\varphi\sin\delta)^{2}\\
                           &+\;\;\;d_{33}\sin^{3}\theta\cos^{2}\delta\sin\delta
    \end{split}
\end{equation}
and
\begin{equation}\label{eq: deff-fsf}
    \begin{split}
d_{\text{eff}}^{\,fsf}&=d_{\text{eff}}^{\,sff} \\
                      &=\;-\;d_{15}\,[\sin\theta \cos\delta\, (\cos\theta \sin\varphi \sin\delta -\cos\varphi \cos\delta )^{2}\\
                      &+\sin\theta \sin\delta\, (\cos\theta \sin\varphi \sin\delta -\cos\varphi \cos\delta )\\
                      &\times (\cos\theta \sin\varphi \cos\delta +\cos\varphi \sin\delta)]\\
                      &-\;d_{24}\,[\sin\theta \cos\delta\, (\cos\theta \sin\varphi \sin\delta +\sin\varphi \cos\delta)^{2}\\
                      &+\sin\theta \sin\delta\, (\cos\theta \cos\varphi \cos\delta -\sin\varphi \sin\delta)\\
                      &\times (\cos\theta \cos\varphi \sin\delta+\sin\varphi \cos\delta)]\\
                      &-\;d_{31}\,\sin\theta \sin\delta\, (\cos\theta \sin\varphi \sin\delta -\cos\varphi \cos\delta )\\
                      &\times (\cos\theta \sin\varphi \cos\delta +\cos\varphi \sin\delta)\\
                      &-\;d_{32}\,\sin\theta \sin\delta\, (\cos\theta \cos\varphi \cos\delta -\sin\varphi \sin\delta)\\
                      &\times (\cos\theta \cos\varphi \sin\delta +\sin\varphi \cos\delta )\\
                      &-\;d_{33}\,\sin^{3}\theta \sin^{2}\delta \cos\delta
    \end{split}
\end{equation}
for type-I and type-II phase-matching, respectively, where the
superscripts "$s$" and "$f$" stand for the "slow" and "fast"
eigenmodes of polarization and their sequence follows the usual
convention $\lambda_{1}\lambda_{2}\lambda_{3}$ with
$\lambda_{1}\geq \lambda_{2}> \lambda_{3}$. The angle $\delta$
whose introduction simplifies the expressions and which is
determined from
\begin{equation} \label{eq: delta-angle}
    \tan2\delta =\frac{\cos\theta \sin2\varphi}{\cot^{2}V_{Z}\sin^{2}\theta +\sin^{2}\varphi -\cos^{2}\theta \cos^{2}\varphi}
\end{equation}
$(0<2\delta <\pi)$ defines the polarization directions of the slow
and fast waves which are orthogonal to each other. It is the angle
between the polarization direction of the slow wave and the plane
of propagation containing the principal Z-axis. An approach for
the generalization of the $d_{\text{eff}}$ expressions in the case
of collinear interactions taking into account the spatial walk-off
can be found in Ref.~\cite{diesperov}.

In the principal planes the expressions for $d_{\text{eff}}$
(\ref{eq: deff-ssf})-(\ref{eq: deff-fsf}) are reduced to:
\begin{eqnarray}
    d_{\text{eff}}^{\,eoe}&=& d_{\text{eff}}^{\,oee}\nonumber\\
    &=& -(d_{24}\,\sin^{2}\varphi + d_{15}\,\cos^{2}\varphi) \,(\text{X-Y plane}) \label{eq: deff-XY}\\
    d_{\text{eff}}^{\,oeo}&=&d_{\text{eff}}^{\,eoo}=-d_{24}\,\sin\theta \quad (\text{Y-Z plane}) \label{eq: deff-YZ}\\
    d_{\text{eff}}^{\,ooe}&=&+d_{31}\,\sin\theta \quad (\text{X-Z plane}, \theta < V_{Z}) \label{eq: deff-XZm}\\
    d_{\text{eff}}^{\,oeo}&=&d_{\text{eff}}^{\,eoo}=-d_{15}\,\sin\theta \; (\text{X-Z plane}, \theta > V_{Z}) \label{eq: deff-XZp}
\end{eqnarray}
with superscripts "$o$" and "$e$" denoting the ordinary and
extraordinary beams. LIS behaves as an optically negative uniaxial
crystal in the X-Y and X-Z (for $\theta <V_{Z}$) planes and as an
optically positive uniaxial crystal in the Y-Z and X-Z (for
$\theta >V_{Z}$) planes. Assuming the Kleinman symmetry condition
to hold then $d_{15}=d_{31}$ and $d_{24}=d_{32}$.

The phase-matching loci for collinear SHG can be categorized using
the classification of Hobden~\cite{hobden}. Our calculations based
on the relatively simple transcendental equations derived by Yao
\emph{et al}~\cite{yao}, predict that no phase-matched SHG is
possible below 1617 nm. LIS enters Hobden class 13 at 1617 nm,
class 11 at 1783 nm, class 10 at 2353 nm, and class 9 at 2675 nm,
then it goes back to class 10 at 5493 nm, to class 11 at 6111 nm,
to class 13 at 8224 nm, and finally no SHG phase-matching is
possible again above 8710 nm. These transitional fundamental
wavelengths correspond to propagation directions along the Y - or
along the Z - principal axes (noncritical phase-matching) either
for the $ssf$ or the $fsf$ ($sff$) polarization configurations.

\begin{figure}[tb]
    \includegraphics[width=6cm]{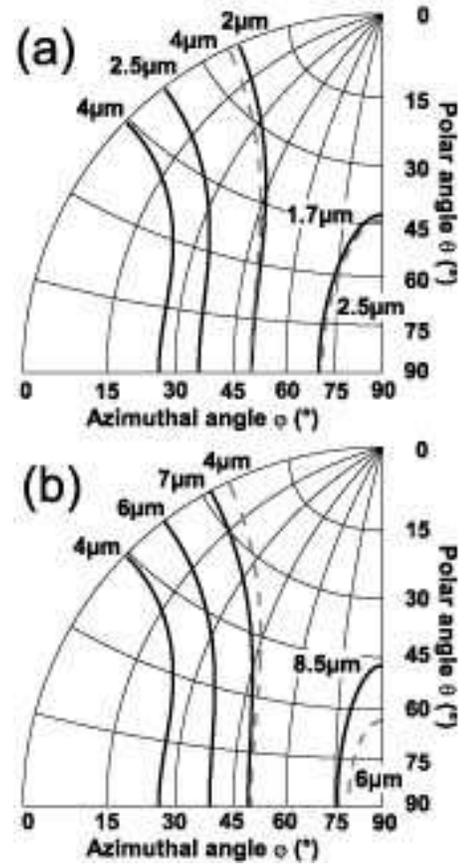}
    \caption{\label{fig: stereograph}  (a)-(b): Stereographic projections of the SHG in the first octant of LIS calculated
     for wavelengths representative of the Hobden classes. Type-I ($ssf$) interaction (solid lines) and type-II ($fsf$,
     $sff$) interactions (dashed lines). The wavelength annotations refer to the fundamental ones.}
  \end{figure}

Figs.~\ref{fig: stereograph}a,b illustrate the different classes
for SHG in LIS at several representative fundamental wavelengths.
The surface of the unit sphere is projected onto the X-Z plane of
the crystal. The direction of the wave vectors of the interacting
waves for phase-matching as given by their interception with the
surface of the unit sphere is plotted.
The stereographic projection of the first octant is presented only
but the loci in the other octants can be obtained by mirror
reflections across the principal planes. We do not present plots
of the angle $\delta$ which can be easily calculated from
Eq.(\ref{eq: delta-angle}). Its dispersive properties (determined
by the angle $V_{Z}$) lead to slightly different values at the
three interacting wavelengths but normally the effect on the
conversion efficiency can be neglected. Qualitative arguments
relating the spatial walk-off (magnitude and direction) to the
topology depicted in Fig.~\ref{fig: stereograph} can be found in
Ref.~\cite{roberts}. The calculation of the spatial walk-off for
an arbitrary propagation direction is beyond the scope of this
paper but several works provide suitable approaches to this
aim~\cite{brehat,yao2,zhang,zadorozhnii}. Analytical approaches
for the estimation of acceptance parameters based on the
small-signal approximation in the general case of biaxial crystals
can be found in Ref.~\cite{yao2,zhang03}. We note here only, that
since the angle $V_{Z}$ for LIS is not far from $90^{\circ}$ the
dispersive properties and in particular the spectral acceptance
are not expected to be substantially modified by propagation
outside the principal planes. Fig.~\ref{fig: stereograph}
indicates that double solutions for the angles exist (e.g. for
Hobden's class 9) which is equivalent to the existence of points
outside the principal planes where
$\partial\varphi/\partial\theta=0$ holds. This means that
noncritical phase-matching in one direction can also occur outside
the principal planes (e.g. at
$\varphi=26.7^{\circ},\theta=74.35^{\circ}$ for $ssf$ type SHG at
$4\,\mu$m).

\subsection{\label{subsec: principalSHG} SHG in principal planes}
\begin{figure}[tb]
    \includegraphics[width=\columnwidth]{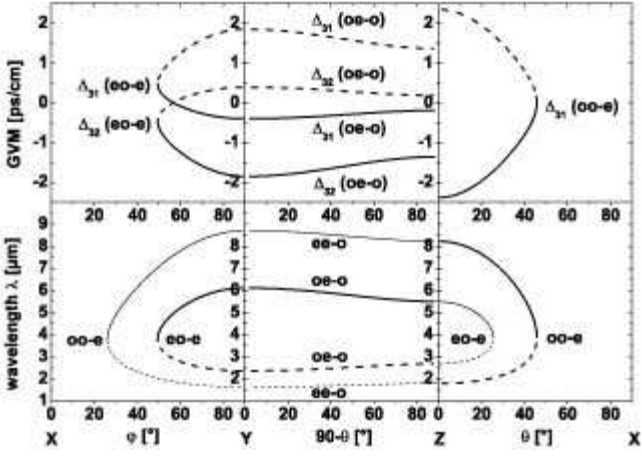}
    \caption{\label{fig: principalSHG} SHG phase-matching in the principal planes of LIS.
    Thick lines in the lower part show fundamental wavelengths for which $d_{\text{eff}}\neq 0$ and thin lines
    indicate cases where $d_{\text{eff}}$ vanishes. The inverse group velocity
    mismatch GVM
     ($\Delta_{31}=v_{3}^{-1}-v_{1}^{-1}$ and  $\Delta_{32}=v_{3}^{-1}-v_{2}^{-1}$ where $v_{1}, v_{2}, v_{3}$ denote
      the group velocities $\partial\omega_{i}/\partial k_{i}$ at $\lambda_{1},\lambda_{2}$, and $\lambda_{3}$) is shown
       in the upper part for the cases where $d_{\text{eff}}\neq 0$. The solid (dashed) lines correspond to the branch
        with longer (shorter) wavelengths.}
  \end{figure}

The SHG phase-matching directions in the principal planes are
shown in the lower part of Fig.~\ref{fig: principalSHG}.  The two
first panels (X-Y) and (Y-Z) in the bottom part correspond
respectively to the experimental phase-matching plots in the main
and inset panels of Fig.~\ref{fig: shg-pm}. The transitional
wavelengths enumerated above can be easily identified there. The
SHG ranges where $d_{\text{eff}}\neq 0$ are 1783-8224 nm for
type-I ($ooe$) phase-matching in the X-Z plane, 2353-6111 nm for
type-II ($eoe$) phase-matching in the X-Y plane, and 2353-2675 nm
and 5493-6111 nm for type-II ($oeo$) phase-matching in the Y-Z
plane. We observe an interesting feature in this crystal: the SHG
limits with $d_{\text{eff}}\neq 0$ are larger for propagation
\emph{outside} the principal planes where type-I phase-matching
down to 1617 nm and up to 8710 nm is possible. Under the
convention $n_{X}<n_{Y}<n_{Z}$ the largest birefringence and
consequently the shortest SHG wavelength is obviously achieved for
type-I interaction and propagation along the Y-axis. However, in
LIS when the Y-axis is approached in the principal planes X-Y or
Y-Z $d_{\text{eff}}$ for type-I interaction vanishes and this is
true also for the limiting case of propagation along the Y-axis.
This is the reason why propagation outside the principal planes
can be used e.g. to shorten the SHG lower wavelength limit
(Fig.~\ref{fig: stereograph}). A similar situation is known to
exist and has been experimentally verified in KTP~\cite{mooren}.

\begin{figure}[tb]
    \includegraphics[width=\columnwidth]{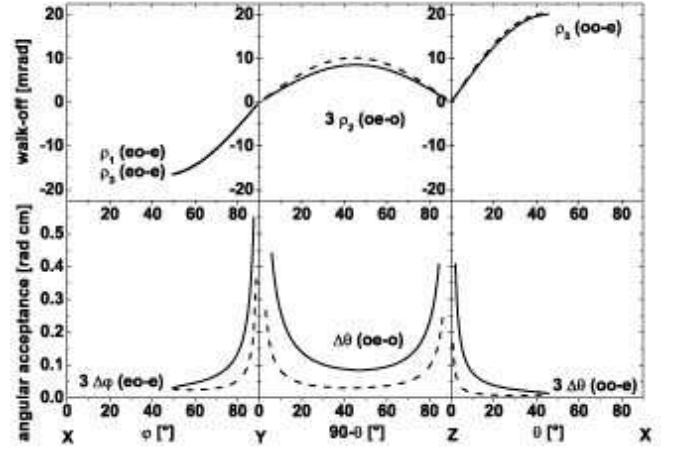}
    \caption{\label{fig: accept-wo} SHG internal angular acceptance (bottom) and walk-off angles (top) in the principal
    planes of LIS. Thick solid lines correspond to the branches with longer wavelengths from Fig.~\ref{fig: principalSHG}
    and thick dashed lines correspond to the branches with shorter wavelengths. Only the cases with $d_{\text{eff}}\neq 0$ are included.}
  \end{figure}

It is seen from Fig.~\ref{fig: principalSHG} that in the Y-Z plane
the type-II interaction is quasi angle-noncritical (but
wavelength-critical, see inset of Fig.~\ref{fig: shg-pm}) which
ensures a large acceptance angle and a small walk-off angle. In
contrast, type-I interaction in the X-Z plane and type-II
interaction in the X-Y plane have regions of
quasi-wavelength-noncritical phase-matching
($\partial\Phi/\partial\lambda\sim 0$) centered at 3889 nm and
3803 nm, respectively (Fig.~\ref{fig: principalSHG}, see also
Fig.~\ref{fig: shg-pm}). Letting
$\Phi=\varphi\;\text{or}\;\theta$, the walk-off angles in the
upper panels of Fig.~\ref{fig: accept-wo} are calculated using the
simplified formula $\tan\rho_{i}=[n_i^{e}(\Phi)]^{-1}\partial
n_i^{e}/\partial\Phi$ valid for uniaxial crystals, where the
subscript $i=1,2,3$ is associated with wavelength $\lambda_i$
($\lambda_3^{-1}=\lambda_2^{-1}+\lambda_1^{-1}$) and
$n_i^{e}(\Phi)$ is the extraordinary index of refraction of the
walking-off wave as given by the uniaxial analogy in the principal
planes. We preserved its sign which is in accordance with the
corresponding phase-matching angles i.e. a positive value of the
walk-off means that the Poynting vector is at an angle larger than
the phase-matching angle and vice-versa.
 Note that the walk-off is
very similar for the two branches of the solution for
phase-matching in all three principal planes. In the X-Y plane
these branches and the two different walk-off parameters (we have
here two extraordinary waves) are almost undistinguishable. The
maximum walk-off amounts to $\rho_{3}=+1.15^{\circ}$ in the X-Z
plane for SHG at 3889 nm.

\begin{figure}[tb]
    \includegraphics[width=8cm]{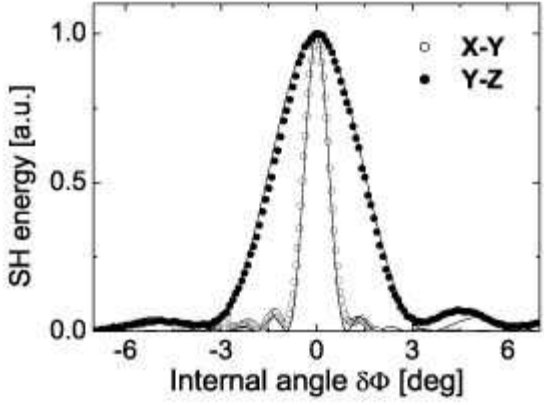}
    \caption{\label{fig: comp-xy-yz} Experimental angular tuning curves for type-II SHG of $\lambda=2590\,$nm
    in the X-Y plane (at $\Phi_{\text{PM}}\equiv\varphi_{\text{PM}}=66.2^{\circ}$, blank circles) and in the Y-Z plane (
    at $\Phi_{\text{PM}}\equiv\theta_{\text{PM}}=27.9^{\circ}$,
    black circles). The larger acceptance angle in the Y-Z
    plane is due to the much smaller
    $\Delta n_{\text{Y-Z}}=-0.007$ birefringence as compared with $\Delta
    n_{\text{X-Y}}=-0.035$ (Fig.~\ref{fig: indices}). The walk-off angles are
    $\rho_{1}\simeq
    \rho_{3}=12.2\,$mrad (X-Y), $\rho_{2}=2.8\,$mrad (Y-Z). The crystal lengths are
    $L=6\,$mm (sample LIS(1) used in X-Y) and $L=7\,$mm (sample LIS(2) used in Y-Z plane), see Subsection~\ref{subsec: gaussian-beam}.
    The fundamental beam (diameter $2w_0\approx 2.6\,$mm) is the idler output of a ns Nd:YAG-pumped LiNbO$_{3}$ OPO. }
  \end{figure}

The acceptance angle is evaluated from the phase velocity mismatch
$\Delta k(\delta\Phi)$ due to an angular deviation
$\delta\Phi=\Phi-\Phi_{\text{PM}}$ of the wavevectors around the
nominal phase-matched direction $\Phi_{\text{PM}}$. A Taylor
expansion of $\Delta k$ leads to
\begin{equation} \label{eq: deltak}
    \Delta
    k(\delta\Phi)=\gamma_{\text{CPM}}\,\delta\Phi+\gamma_{\text{NCPM}}(\delta\Phi)^{2}+...
\end{equation}
where $\gamma_{\text{CPM}}$=$\left[\partial(\Delta
k)/\partial\Phi\right]_{\Phi=\Phi_{\text{PM}}}$ vanishes for
non-critical phase-matching (NCPM, i.e. for $\Phi=0,90^{\circ}$)
and $\gamma_{\text{NCPM}}$=$\frac{1}{2}\left[\partial^{2}(\Delta
k)/\partial\Phi^{2}\right]_{\Phi=\Phi_{\text{PM}}}$. The
acceptance angle, defined as the bandwidth at FWHM of the
$\text{sinc}^{2}(\Delta k L/2)$ phase-mismatch function, is given
by $\Delta\Phi=2.784/|\gamma_{\text{CPM}}|L$ for a critical
phase-matching (CPM), while for NCPM it is
$\Delta\Phi=|2.784/\gamma_{\text{NCPM}}L|^{1/2}$. The lower panels
of Fig.~\ref{fig: accept-wo} display the acceptance angles (for a
length $L=1\,$cm) in the three principal planes. The acceptance
curves, computed with $\gamma_{\text{CPM}}$ only in the expansion
(\ref{eq: deltak}), are interrupted near non-criticality at
maximum values corresponding to the ones calculated for $L=1\,$cm
using $\gamma_{\text{NCPM}}$ only. Although that type of
presentation in Fig.~\ref{fig: accept-wo} is not very accurate in
these limits we still prefer it since it permits the results to be
presented as scalable with respect to the crystal length $L$
(simultaneous consideration of both derivatives would not allow
this). The experimental comparison of type-II SHG acceptance
angles in the Y-Z and X-Y planes is shown with symbols in
Fig.~\ref{fig: comp-xy-yz}, for $\lambda=2590\,$nm. The solid
lines depict the plane-wave $\text{sinc}^{2}(\Delta k\,L/2)$
phase-mismatch tuning functions computed with the first-order term
in Eq.(\ref{eq: deltak}), with
$\gamma_{\text{CPM}}(\theta_{\text{PM}})=(\pi/\lambda)[n_{2}^e\rho_{2}]_{\theta_{\text{PM}}}$=$7.2\times
10^{-3}\mu$m$^{-1}$ (type-II ($oeo$) SHG in the Y-Z plane) and
$\gamma_{\text{CPM}}(\varphi_{\text{PM}})=(\pi/\lambda)[2n_{3}^{e}\rho_{3}-n_{1}^{e}\rho_{1}]_{\varphi_{\text{PM}}}$=
$3.17\times 10^{-2}\mu$m$^{-1}$ (type-II ($eoe$) SHG in the X-Y
plane). The perfect match of the theoretical curves with the
experimental symbols highlights both the accuracy of the
dispersion data and the single-domain feature of the samples used.

The chosen presentation of the inverse group velocity mismatch in
the upper part of Fig.~\ref{fig: principalSHG} is equivalent to
the spectral acceptance $\Delta\nu$ but contains the sign as an
additional information. In the simplest cases of type-I SHG or
degenerate DFG we have e.g. $\Delta\nu L=0.886/|\Delta_{31}|$ and
the gain bandwidth in optical parametric amplification assuming a
narrow-band pump wave at $\lambda_{3}$ is inversely proportional
to $|\Delta_{21}| =| 1/v_{2}-1/v_{1}|$. The two parameters
$\Delta_{31}$ and $\Delta_{32}$ vanish at 3889 nm in the X-Z plane
which means large spectral acceptance for the SHG of short pulses
where the second derivative of the wavevector-mismatch comes into
play. The situation is different for type-II interaction in the
X-Y plane: here $\Delta_{32}$ = -$\Delta_{31}$ at 3803 nm, at 2915
nm $\Delta_{32}$ vanishes and at 4946 nm $\Delta_{31}$ vanishes
but in none of these cases an extremum of the spectral acceptance
occurs because in type-II SHG of short pulses all three waves
should be considered as broad-band. Thus long crystals can be used
to increase the conversion efficiency for type-I SHG in the X-Z
plane near 3889 nm even at femtosecond pulse durations. For
type-II SHG of femtosecond pulses near 3803 nm in the X-Y plane
the second harmonic will not be lengthened either because the
temporal walk-off between the two polarizations of the fundamental
wave will limit the interaction length, but consequently the
conversion efficiency will be low.

\subsection{\label{subsec: sum-and-diff} Sum-and-difference frequency mixing}
Type-I phase-matching in the X-Z plane for non-degenerate
three-wave interactions (sum- and difference-frequency mixing as
well as optical parametric generation, amplification and
oscillation) is presented in Fig.~\ref{fig: opaxz} where two
branches of the solution can be seen. The whole transparency range
of LIS can be covered for $0^{\circ}<\theta<40^{\circ}$ in the X-Z
plane but $d_{\text{eff}}$ increases with the phase-matching angle
$\theta$. Note that $d_{\text{eff}}=0$ for NCPM
($\theta=0^{\circ}$) and that the situation $\theta>V_{Z}$
(Eq.~\ref{eq: deff-XZp}) is never reached. At larger angles
$\theta$ we observe a retracing behavior in the left branch: e.g.
in the case of OPO, one and the same pump wavelength $\lambda_{3}$
corresponds to two pairs $(\lambda_{1},\lambda_{2})$ of
signal-idler wavelengths (curve 5). In this region the spectral
acceptance is very large. Thus for $\theta=40^{\circ}$ (left
branch of curve 5), $\lambda_{3}=900\,$nm and
$\lambda_{2}=1150\,$nm all three group velocities are very close:
at the point where $\Delta_{31}=0$ we have
$\Delta_{32}=23\,$fs/mm. This means that this phase-matching
configuration is especially suitable for frequency conversion of
femtosecond pulses. In optical parametric amplifiers (OPA's) this
advantage can be utilized, however, only in combination with a
control of the spectral bandwidth through the seed signal. In the
regions near the degeneracy points (SHG points) we have on the
other hand $\Delta_{21}= 0$ but the wave at $\lambda_{3}$ can have
in general a different group velocity. Such a regime is attractive
for broadband parametric amplification in the field of a
narrow-band pump pulse as in the case of chirped pulse optical
parametric amplification~\cite{rotermund2002}. For
femtosecond-laser based frequency metrology
applications~\cite{fs-comb}, this phase-matching when implemented
in a synchronously-pumped OPO would provide a wide frequency comb
grid spanning the $10-12\,\mu$m region useful for high precision
spectroscopy of spherical reference standard molecules. Increasing
the phase-matching angle (curve 6 for $\theta=45^{\circ}$ in
Fig.~\ref{fig: opaxz}) the two branches merge into a closed
contour and we approach the point where SHG phase-matching only
for a single wavelength is possible (see Fig.~\ref{fig:
principalSHG}) and all three group velocities are again very close
but the tunability in that case is very limited.

\begin{figure}[t]
    \includegraphics[width=7cm]{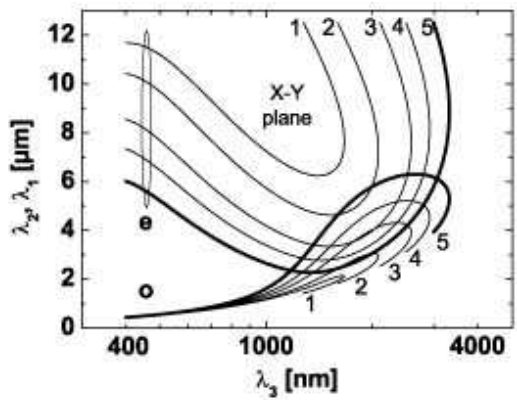}
    \caption{\label{fig: opaxy} Type-II ($eoe$ and $oee$) phase-matching for sum- and difference-frequency generation
    in the X-Y plane of LIS and several values of the azimuthal angle $\varphi$ : $35^{\circ}$ (lines 1), $40^{\circ}$
    (lines 2), $50^{\circ}$ (lines 3), $60^{\circ}$ (lines 4) and $90^{\circ}$- NCPM (thick lines 5).
    The curves are terminated at the left and top side by the transparency range of the crystal. }
  \end{figure}

The curves for type-II phase-matching in the X-Y plane
(Fig.~\ref{fig: opaxy}) have a completely different shape. The two
branches of the solution are represented by curves of opposite
curvature which can cross at two points where phase-matching for
degenerate DFG or SHG occurs. With decreasing phase-matching angle
these branches separate and a single crossing point is reached
(see the curves 3 for $\varphi=50^{\circ}$) which corresponds to
the single SHG solution in Fig.~\ref{fig: principalSHG}. For yet
smaller angles no crossing occurs and the degeneracy point is not
reached. For all phase-matching curves presented we observe again
a retracing behavior at longer $\lambda_{3}$: for each
$\lambda_{3}$ two couples $(\lambda_{1},\lambda_{2})$ are
phase-matched. At the point where these two pairs merge into one
(at the maximum $\lambda_{3}$ permitting phase-matching) the waves
at $\lambda_{1}$ and $\lambda_{2}$ have equal group velocities and
similarly to the case discussed for Fig.~\ref{fig: opaxz}
broadband parametric amplification with a narrow-band pump wave
can be realized. The deviation from the pump group velocity
remains, however, essential. Thus e.g. at $\varphi=90^{\circ}$ and
$\lambda_{3}=3310\,$nm, $\Delta_{21}\simeq 0$ and
$\Delta_{31}\simeq \Delta_{32}\simeq -1.65\,$ps/cm. Improved pump
group-matching at this point occurs when decreasing the
phase-matching angle $\varphi$ in accordance with Fig.~\ref{fig:
principalSHG}.

Comparing Fig.~\ref{fig: opaxy} to Fig.~\ref{fig: opaxz} we note
that at relatively short $\lambda_{3}$ (e. g. 1064 nm) full
tunability with a fixed crystal cut is achievable only in the X-Z
principal plane. The group velocity mismatch depends on the
specific wavelengths chosen. At $\lambda_{3}=1064\,$nm (a case
interesting for parametric down-conversion)
$\Delta_{21}(X-Z)<\Delta_{21}(X-Y)$ up to $\lambda_{1}= 10\,\mu$m.
This means that in this practical wavelength range type-II
phase-matching in the X-Y plane is more advantageous for the
development of narrow-band parametric generators or oscillators.

Our analysis of the phase-matching properties of LIS is just a
first order approximation because it was based on the small signal
limit where saturation is neglected. We considered also only
collinear interactions. Non-collinear interactions attracted
recently great interest because in addition to the possibility to
improve the tunability, to spatially separate the beams, and to
compensate for the spatial walk-off, this can considerably reduce
the group velocity mismatch~\cite{rotermund2001}. The latter is
very promising for short pulse interactions. Explicit expressions
for non-collinear phase-matching in the principal planes of a
biaxial crystal can be found in Ref.~\cite{liu2001} and
calculations of non-collinear group velocity matching for
arbitrary wave vector directions in Ref.~\cite{zhang03}.

\section{\label{sec: meas-deff} Second-order nonlinear coefficients}

There has been only one characterization study of the nonlinear
properties of LIS made by Boyd \emph{et al} in the early
seventies~\cite{boyd}. The samples used then were of limited size
and poor quality, and besides their results were derived only from
non-phase matched SHG at $10.6\,\mu$m (wedge technique). Further
their measurements were performed relative to the nonlinearity of
GaAs, that was several times corrected during the past years. So
it is important to confirm or to correct these previous results
using phase-matched second-order interactions on mature samples of
adequate size.

The elements $d_{ij}$ of the nonlinear tensor (\ref{eq: d-tensor})
have been determined from the measurement of the SHG efficiency of
various laser sources, using either type-I or type-II interactions
in the principal planes of LIS and also out of them. Two kinds of
measurements have been performed: in one kind (Subsection A), the
source is the idler output (in the range $2.4-2.6\,\mu$m) of a
12-ns pulsewidth, 10-Hz repetition rate, type-I
LiNbO$_3$-OPO~\cite{assl}. Long samples were used, and hence the
residual absorption and the beam walk-off effect, as well as the
focusing effects, have to be taken into account to derive the
nonlinear coefficients. The second independent series of SHG
experiments (Subsection B) used femtosecond pulses and very thin
oriented samples. This allows to neglect the effect of absorption
and beam walk-off and focusing effects, the data analysis in the
low-depletion limit being derived from the standard plane-wave SHG
theory.

\subsection{\label{subsec: gaussian-beam} Measurements based on Gaussian beam SHG theory}
In order to remain in the low pump depletion limit, the peak
on-axis fundamental intensity used was below 10 MW/cm$^{2}$, and
the maximum idler energy of the LiNbO$_3$-OPO was 10 mJ at
$\lambda=2590\,$nm. The spectral linewidth of the idler was below
15 GHz, and the transverse beam profile ressembled a Gaussian.
Three thick samples of different coloration (rose and yellowish)
were used in conjunction with weak focusing producing Gaussian
beam waist of either $w_{0}\sim 1\,$mm or $w_{0}\sim 1.3\,$mm.
Sample LIS(1) was rose-annealed, cut at $(\varphi=66.5^{\circ},
\theta=90^{\circ})$ for type-II ($eoe$) SHG in the X-Y plane, with
dimensions $4\times 4\times 6\,$mm$^3$ ($L=6\,$mm). Sample LIS(2)
was yellowish (as-grown), cut at $(\varphi=90^{\circ},
\theta=28^{\circ})$ for type-II ($oeo$) SHG in the Y-Z plane, with
dimensions $4\times 4\times 7\,$mm$^3$ ($L=7\,$mm). LIS(1) and
LIS(2) were the same samples already used in
Subsection~\ref{subsec: sellmeier}, see Fig.~\ref{fig: shg-pm} and
also Fig.~\ref{fig: comp-xy-yz}. Sample LIS(3) had almost
identical cut as LIS(1) $(\varphi=66^{\circ}, \theta=90^{\circ})$,
but was of yellowish tinge and with dimensions $4\times 4\times
5\,$mm$^3$ ($L=5\,$mm). We have checked that, as the linear
dispersion data, the nonlinear coefficients do not significantly
depend on coloration when the additional absorption effect of the
yellow samples was taken into account. For typical pump energy
$E_{\omega}=1\,$mJ and $w_{0}=1.3\,$mm, LIS(2) yielded
$E_{2\omega}=2.5\,\mu$J, while LIS(1) yielded
$E_{2\omega}=11\,\mu$J and LIS(3), $E_{2\omega}=7\,\mu$J. The
difference between the efficiencies of LIS(1) and LIS(3) was
mainly due to the much stronger second harmonic absorption of the
latter ($\alpha_{2\omega}=0.27\,$cm$^{-1}$)~\cite{photonicwest}.
For LIS(1), the residual fundamental and harmonic absorption
amounted to $\alpha_{\omega,2\omega}=0.05-0.06\,$cm$^{-1}$ while
for LIS(2) $\alpha_{\omega,2\omega}\simeq
0.6\,$cm$^{-1}$~[\cite{photonicwest}].

The measurements were performed relatively to a dual-band
anti-reflection-coated KTP sample with $L=5.93\,$mm, cut at
$(\varphi=0^{\circ},\theta=56.4^{\circ})$ for SHG of
$\lambda=2.53\,\mu$m in the X-Z plane. The measured nonlinear
coefficient of KTP was $d_{24}=2.3\pm
0.2\,$pm/V~\cite{type-II-ktpshg} and served as a reference,
without having to correct for dispersion since the measurement
wavelength is almost identical to that used for LIS. We only
corrected for the residual absorption loss
($\alpha_{\omega}=0.068\,$cm$^{-1}$ at $2.59\,\mu$m) of this KTP
sample, due to OH$^{-}$ band~\cite{type-II-ktpshg}. According to
the Gaussian beam SHG theory in the nanosecond pulse regime, the
effective nonlinearity in the absence of depletion scales
as~\cite{eckardt}
\begin{equation} \label{eq: deff-eckardt}
    d_{\text{eff}}^{\,2}=\Gamma_{\text{SH}}
    \frac{\tau_\omega^2}{\tau_{2\omega}} \frac{c\epsilon_0 n^2 \lambda^3}{16\pi^2 Lh}
\end{equation}
where $\Gamma_{\text{SH}}=E_{2\omega}/E_\omega^2$ is the pulse
energy conversion efficiency (in unit of J$^{-1}$), $n$ is the
average index of refraction at fundamental and harmonic
wavelengths and $h$ is the walk-off-focusing function that
accounts for diffraction, absorption and beam-walk-off effects.
The exact expression of $h$ for type-II interaction can be found
in Ref.~\cite{zondycomp}. The second harmonic pulse duration was
taken as $\tau_{2\omega}\simeq \tau_{\omega}/\sqrt{2}$. In the
nanosecond regime and for the given crystal lengths,
group-velocity mismatch can be safely neglected in SHG. The
advantage of measuring the LIS coefficient relative to a reference
material stems from the fact that it is neither necessary to
evaluate the absolute energies nor to take into account the loss
due to the numerous filter sets used to isolate the fundamental
and second harmonic field, since from Eq.(\ref{eq: deff-eckardt}),
the ratio of the two $d_{\text{eff}}$'s scales as
\begin{equation} \label{eq: deff-ratio}
    \frac{d_{\text{eff}}^{\,2}(\text{LIS})}{d_{\text{eff}}^{\,2}(\text{KTP})}=\frac{\Gamma_{\text{SH}}(\text{LIS})}{\Gamma_{\text{SH}}(\text{KTP})}
    \frac{n^2(\text{LIS})}{n^2(\text{KTP})}
    \frac{L(\text{KTP})}{L(\text{LIS})}\frac{h(\text{KTP})}{h(\text{LIS})}.
\end{equation}
The efficiencies $\Gamma_{\text{SH}}$ were derived from the linear
fit of the plot of $E_{2\omega}$ versus $E_{\omega}^2$, for
fundamental energies ranging from 0 to 5 mJ~\cite{photonicwest}.
The results of the data analysis, averaged over the two waist
values, are the following:
\begin{eqnarray}
    d_{\text{eff}}[\text{LIS}(2)]= 1.36\;(\pm 15\%)\; d_{\text{eff}}(\text{KTP}), \label{eq: deff-ratio-num1}\\
    d_{\text{eff}}[\text{LIS}(1)]= 3.05\;(\pm 15\%)\; d_{\text{eff}}(\text{KTP}), \label{eq: deff-ratio-num2}\\
    d_{\text{eff}}[\text{LIS}(3)]= 3.15\;(\pm 15\%)\; d_{\text{eff}}(\text{KTP}). \label{eq: deff-ratio-num3}
\end{eqnarray}
The rather large uncertainties are mainly due to the uncertainty
in the absorption coefficients through the evaluation of the $h$
focusing functions. Given that
$d_{\text{eff}}(\text{KTP})=d_{24}\,\sin\theta$
($\theta=58.7^\circ$ at $\lambda=2590\,$nm), from (\ref{eq:
deff-ratio-num1}) and (\ref{eq: deff-YZ}) one derives
$d_{24}(\text{LIS})= 5.73\;\pm 20\%\,\text{pm/V}$. The two last
relations (\ref{eq: deff-ratio-num2})-(\ref{eq: deff-ratio-num3})
were averaged to yield, using (\ref{eq: deff-XY}) and the
determined value of $d_{24}$, $d_{15}(\text{LIS})= 7.94\;\pm
20\%\,\text{pm/V}$. The ratio of the two nonlinear coefficients is
hence found to be $d_{24}/d_{15}=0.72$.

\subsection{\label{subsec: femto-beam} Measurements based on femtosecond SHG}
The SHG experiments with femtosecond pulses were performed also in
the low depletion limit for the fundamental in order to avoid
complications from saturation effects and spatial effects across
the beam cross section. The coefficient $d_{31}$ was estimated by
type-I SHG in the X-Z plane from Eq.(\ref{eq: deff-XZm}), once it
was known it was used for the determination of $d_{24}$ from
type-II SHG in the X-Y plane using Eq.(\ref{eq: deff-XY}) and
assuming Kleinman symmetry to hold (i.e. $d_{15}=d_{31}$). Finally
the diagonal element $d_{33}$ was measured by type-I SHG outside
the principal planes using Eq.(\ref{eq: deff-ssf}) and the already
determined off-diagonal components of the $\mathbf{d}^{(2)}$
-tensor, again assuming Kleinman symmetry. Note that a similar
strategy was used previously to determine $d_{33}$ of KTP by
phase-matched SHG~\cite{boulanger-sphere} and as outlined there
the error in the determination of this diagonal element is
normally larger since it accumulates the uncertainties of the
other coefficients. Alternatively the plane Y-Z could be used for
determination of $d_{24}$ as done in the previous subsection,
however, the spectral tunability is very narrow in this principal
plane (see Fig.~\ref{fig: principalSHG}) which is not adequate for
measurements with broadband femtosecond pulses.

All three LIS samples used here were annealed and rose in colour,
had an aperture of $4\times 5\,$mm$^2$ and were $0.2\,$mm thick.
They were cut at ($\varphi=0^{\circ},\theta=34^{\circ}$) [LIS(4)],
($\varphi=59^{\circ},\theta=90^{\circ}$) [LIS(5)], and
($\varphi=30^{\circ},\theta=44^{\circ}$) [LIS(6)], respectively.
The small thickness allows to neglect absorption or scattering
losses, the focusing effects and also spatial beam walk-off
effects. This measurement was also relative and we used as a
reference sample an AGS crystal with the same dimensions cut at
$\varphi=45^\circ$ and  $\theta=45^\circ$ for type-I SHG. In order
to minimize the error originating from the still different
group-velocity mismatch the pulsewidth at the fundamental was
chosen relatively large ($160\,$fs in all cases) since LIS samples
with thickness less that 0.2mm could not be prepared with
sufficient quality. This pulsewidth corresponds to spectral
bandwidths of 80-90nm. For $d_{36}$ of AGS we used the value of
$13.9\,$ pm/V~\cite{zondy-ags} which was measured at a similar
wavelength ($\lambda=2.53\,\mu$m) and actually relative to
$d_{24}$ of KTP using the same value for the latter as in the
previous subsection of the present work. In the plane wave
approximation and having in mind the equal fundamental energy and
crystal thickness, $d_{\text{eff}}$ of LIS was determined simply
from
\begin{equation} \label{eq: deff-femto}
    \frac{d_{\text{eff}}^2(\text{LIS})}{d_{\text{eff}}^2(\text{AGS})}=\frac{E_{2\omega}(\text{LIS})}{E_{2\omega}(\text{AGS})}
    \frac{[1-R(\text{AGS})]^3}{[1-R(\text{LIS})]^3}\frac{n^3(\text{LIS})}{n^3(\text{AGS})}
\end{equation}
where $n^3$ is the product of the the three refractive indices
involved, and $R$ takes into account the Fresnel reflections at
the entrance surface (twice for the fundamental) and the exit
surface (second harmonic).

The femtosecond source at 1-kHz repetition rate was a KTP-based
OPA which was seeded by the frequency-doubled idler of a BBO-based
OPA, both OPAs being pumped by the same $800\,$nm, $40\,$fs pump
source (Ti:sapphire regenerative amplifier). The 3-mm KTP used in
the OPA ensured sufficiently long wavelengths that cannot be
achieved by BBO. Seeding by the frequency doubled idler of the
BBO-OPA was preferred against seeding by the signal wavelength for
two reasons: the signal wavelengths available from the BBO-OPA
were not short enough to produce idler wavelengths above
$2.5\,\mu$m with the KTP-OPA, and frequency doubling in a 2-mm
thick BBO crystal of the idler used for seeding resulted in
temporal broadening and spectral narrowing which was a
prerequisite to improve the accuracy. SHG was performed at
$2330\,$nm with LIS(4), $2850\,$nm with LIS(5) and $2300\,$nm for
LIS(6). In all cases more than $5\,\mu$J at the fundamental were
available so that it was possible to reliably measure the SHG with
a large area pyroelectric detector even at low conversion
efficiency. We established that up to a conversion efficiency of
20\% (energy) into the second harmonic the result did not change.
In the case of LIS(6) the polarization angle $\delta$ was adjusted
by a He-Ne laser monitoring the polarization rotation by the
leakage through an analyser till an eigenmode was reached. No
additional improvement of the SHG efficiency could be observed
later by further alignment of this angle which was in good
agreement with the calculated value from Eq.(\ref{eq:
delta-angle}). Note that the dispersion of the optic axis angle
$V_Z$ (see Eq.(\ref{eq: optic-axis}) and Fig.~\ref{fig:
optic-axis}) is relatively small and the deviation between visible
($633\,$nm) and $2300\,$nm is less than $2^\circ$, so that the
error originating from that effect is estimated to be less than
5\%. The following results are values of the $\mathbf{d}^{(2)}$
tensor components scaled to $\lambda=2300\,$nm by using Miller's
rule, i.e. assuming the same wavelength dispersion for the
second-order nonlinearity as for the linear susceptibility,
\begin{eqnarray}
    d_{31}(\text{LIS})&=& 7.25\;(\pm 5\%)\;\text{pm/V}, \label{eq: d31-fem}\\
    d_{24}(\text{LIS})&=& 5.66\;(\pm 10\%)\;\text{pm/V}, \label{eq: d24-fem}\\
    d_{33}(\text{LIS})&=& -16\;(\pm 25\%)\;\text{pm/V}. \label{eq: d33-fem}
\end{eqnarray}
The uncertainties are related to the group velocity mismatch and
to the relatively large angular acceptance using thin samples. The
above results are in very good agreement with the values obtained
in the previous subsection by a completely different strategy and
we also confirmed the different sign of the diagonal element
$d_{33}$ as established by Boyd \emph{et al}~\cite{boyd}. Note
that if the results from Ref.~\cite{boyd} which were obtained by
non-phase-matched SHG are renormalized using a more recent value
of $d_{14}$(GaAs)=$83\,$pm/V~\cite{roberts} we arrive at
$d_{31}=6.14\,$pm/V, $d_{32}=5.31\,$pm/V and $d_{33}=9.8\,$pm/V.
Having in mind the different wavelength ($10.6\,\mu$m) used in
Ref.~\cite{boyd}, we obtain a good agreement within $\pm20\%$ with
the present work for all three nonlinear coefficients.

\subsection{\label{subsec: effective-nonlinearity} Effective nonlinearity of LIS}
\begin{figure}[t]
    \includegraphics[width=7cm]{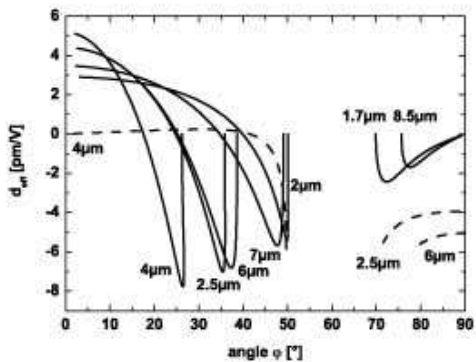}
    \caption{\label{fig: deffoffSHG} Effective nonlinearity for type-I ($ssf$, solid lines) and type-II ($fsf$, dashed lines)
    SHG outside the principal planes of LIS as a function of the azimuthal angle $\varphi$. The labels indicate the fundamental wavelengths
    corresponding to the ($\varphi,\theta$) phase-matching loci of Fig.~\ref{fig: stereograph}.}
  \end{figure}

Once the nonlinear coefficients are determined the effective
nonlinearity $d_{\text{eff}}$ can be easily calculated in the
principal planes using Eqs.~(\ref{eq: deff-XY})-(\ref{eq:
deff-XZm}). The main conclusion drawn in Ref.~\cite{boyd}, namely
that the maximum $d_{\text{eff}}$ is achieved in the X-Y plane
where it is wavelength (angle) independent because of the similar
magnitude of $d_{24}$ and $d_{15}$, remains valid with our updated
values for the $d_{ij}$'s. However, as already mentioned in
subsection~\ref{subsec: principalSHG}, interesting effects can be
expected for propagation outside the principal planes.
Fig.~\ref{fig: deffoffSHG} shows the calculated
$d_{\text{eff}}(\varphi,\theta)$ for SHG using Eq.~(\ref{eq:
deff-ssf})-(\ref{eq: deff-fsf}) with the nonlinearities obtained
in the previous subsection and assuming for simplicity a constant
angle $V_Z=66^\circ$ between the optic axes and the Z-principal
axis (see Fig.~\ref{fig: optic-axis}) which for the wavelengths
selected leads to only negligible errors. The fundamental
wavelengths indicated in the figure correspond exactly to the
phase-matching stereographic projections presented in
Fig.~\ref{fig: stereograph} where the values for the polar angle
$\theta$ for the limiting cases $\varphi=0,90^{\circ}$ can be
seen. As can be seen from the figure for type-I SHG in the Hobden
classes 9, 10, and 11 (see subsection~\ref{subsec: deff} and
Ref.~\cite{hobden}) the improvement in $d_{\text{eff}}$ when
propagating outside the principal planes is not significant. On
the contrary for Hobden class 13, it is definitely advantageous to
use propagation outside the principal planes for maximum
$d_{\text{eff}}$ in type-I SHG. In type-II SHG the curves in
Fig.~\ref{fig: deffoffSHG} have a different character and maximum
$d_{\text{eff}}$ is achieved always in the X-Y plane which is
advantageous as compared to type-II SHG in the Y-Z plane.

\begin{figure}[t]
    \includegraphics[width=7cm]{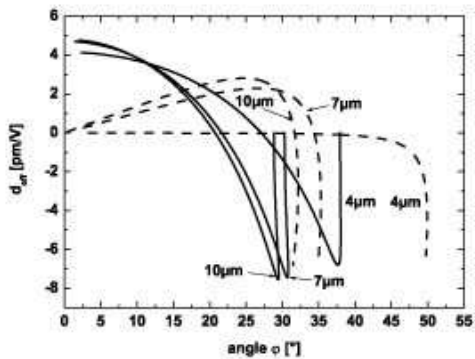}
    \caption{\label{fig: deffoffOPA} Effective nonlinearity for type-I ($ssf$, solid lines) and type-II ($fsf$, dashed lines) down-conversion of
    $\lambda_3=1064\,$nm radiation (optical parametric generation, amplification or oscillation) to three selected idler
    wavelengths $\lambda_1=4, 7, 10\,\mu$m for propagation outside the principal planes of LIS as a function of the azimuthal angle $\varphi$.}
  \end{figure}

For various applications it is important to consider the
down-conversion into the mid-IR of high power sources, e.g. lasers
emitting at 1064nm (Fig.~\ref{fig: deffoffOPA}). For the selected
wavelengths, the absolute negative extremum of the type-II curves
which occurs in X-Y plane (right end, with
$d_{\text{eff}}^{eoe}(90^\circ,\varphi)\simeq -6.5\,$pm/V) is
larger than the extremum of the type-I curves in the X-Z plane
(i.e. on the left side, with
$+4<d_{\text{eff}}^{ooe}(\theta,0^\circ)<5\,$pm/V). Hence type-II
interaction in the X-Y plane has larger $d_{\text{eff}}$ than
type-I interaction in the X-Z plane. However, propagation outside
the principal planes allows to reach a slightly larger maximum
$-7<d_{\text{eff}}^{ssf}(\theta,\varphi)<-7.5\,$pm/V for type-I
interaction which can be advantageous since other parameters like
e.g. the spectral bandwidth can be better in type-I interaction.
We believe the potential for future applications of LIS in the
mid-IR is related also to the utilization of propagation schemes
outside the principal planes.

\section{\label{sec: down-conv} Parametric down-conversion}

\subsection{\label{subsec: opa} Femtosecond optical parametric
amplification} As outlined in section~\ref{sec: introduction} LIS
possesses the largest bandgap among all known mid-IR nonlinear
crystals which results in an extremely low TPA near $800\,$nm.
This unique feature can be utilized for the generation of
femtosecond pulses in the mid-IR by down conversion of the most
widely used Ti:sapphire femtosecond laser systems operating near
$800\,$nm. In Ref.~\cite{rotermund} initial results were presented
in the $4.8-9\,\mu$m region but the gain and the output energy (2
nJ) were extremely low with a $1.5\,$mm thick LIS sample used as a
seeded OPA. Here we report an improvement by a factor of about 50
for the output idler energy and an extension of the tunability
range.

\begin{figure}[t]
    \includegraphics[width=7cm]{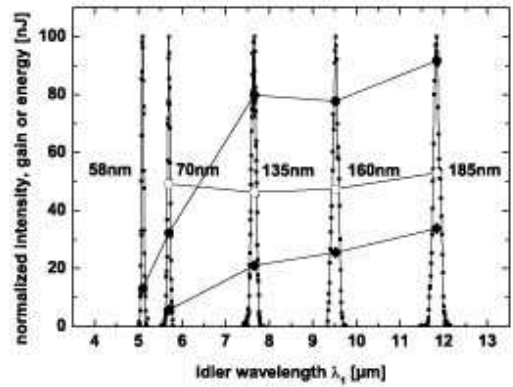}
    \caption{\label{fig: spectra-opa} Idler spectra (the labels indicate FWHM) that demonstrate the achieved tunability with the LIS-based OPA.
    The black circles show the experimentally achieved output idler energy, the blank circles show the achieved (internal) signal gain,
    and the diamonds show the applied seed (signal) energy.}
  \end{figure}

We applied for this experiment a cube of $5\times 5\times
5\,$mm$^3$ of rose annealed LIS cut at $\varphi=41^\circ$ and
$\theta=90^\circ$ for type-II interaction in the X-Y plane. The
single-stage travelling-wave-type OPA was pumped by longer pulses
in order to increase the interaction length. We used $300\,$fs
pump pulses at $820\,$nm from a Ti:sapphire regenerative amplifier
operating at 1kHz. The pulse lengthening was achieved by
adjustment of the pulse compressor and the pump pulses were not
bandwidth-limited (appr. three times the Fourier limit). About
$30\,\mu$J were used to generate continuum in a $2\,$mm thick
sapphire plate and 5 different interference filters between 875
and 975nm were used to select portions (of about 10nm) of the
continuum for seeding the OPA at the signal wavelength
(Fig.~\ref{fig: spectra-opa}). The seed pulse duration was
measured by sum-frequency cross-correlation with a small fraction
of the pump pulses in a $0.5\,$mm thick type-I BBO crystal and
deconvolution yielded a pulse FWHM of the order of $400\,$fs
(about 3 times above the Fourier limit). The OPA was pumped by
$150\,\mu$J and the pump beam parameters corresponded to a peak
on-axis pump intensity of 60 GW/cm$^2$.

The mid-IR spectra depicted in Fig.~\ref{fig: spectra-opa} were
recorded with a $0.5\,$m monochromator ($150\,$gr/mm grating) and
a liquid-N$_2$-cooled HgCdTe detector. Tunability from 5 up to
almost $12\,\mu$m (i.e. the mid-IR cutoff wavelength of
Fig.~\ref{fig: ftir}) could be demonstrated. From 8 to $12\,\mu$m
an almost constant energy level exceeding $80\,$nJ could be
achieved, despite the stronger absorption. This can be attributed
to the lower group velocity mismatch at longer idler
wavelengths~\cite{rotermund} and also to the increasing seed level
which effects compensate for the reduced transmission. The pulse
duration of the idler pulses was estimated from sum-frequency
cross-correlation with a $820\,$nm reference pulse in a $0.3\,$mm
thick type-I AGS crystal in which the group mismatch between the
two input pulses ranged from $715\,$fs/mm at $\lambda_1=9.5\,\mu$m
to $800\,$fs/mm at $\lambda_1=6.5\,\mu$m. Fig.~\ref{fig:
cross-correlation} shows the corresponding cross-correlation
traces. The deconvolved pulse duration at 7.65 and $9.51\,\mu$m
($585\,$fs in both cases) corresponds to bandwidth-limited
Gaussian shaped idler pulses (time bandwidth product equal to
0.44).

Thus LIS is the only nonlinear crystal that permits the direct
frequency conversion of short pulses from the $800\,$nm region to
the mid-IR above $\sim 5\,\mu$m \emph{in a single step} extending
the region covered by crystals like LiNbO$_3$, KNbO$_3$ or KTP and
its isomorphs. The results in this section were obtained with
amplified pump pulses at low (1kHz) repetition rates. Similar
single stage down conversion of high repetition rate ($100\,$MHz)
Ti:sapphire femtosecond oscillators should be possible in
synchronously pumped LIS-based OPOs.
  \begin{figure}[t]
    \includegraphics[width=7cm]{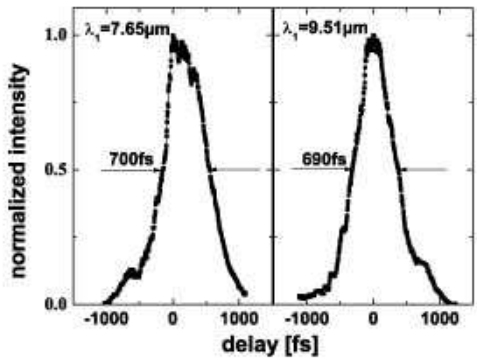}
    \caption{\label{fig: cross-correlation} Cross-correlation traces at two idler wavelengths obtained by sum-frequency mixing
    with a $820\,$nm, $300\,$fs reference pulse. The deconvolved Gaussian pulse durations (FWHM) are in both cases $575\,$fs.}
  \end{figure}

A further potential application of LIS in femtosecond technology
which relies on the extremely low TPA near 800nm~\cite{rotermund}
is related to the generation of single (or sub-) cycle femtosecond
pulses (both at low and high repetition rates) in the mid-IR by
DFG of spectral components belonging to the same broadband
spectrum of an ultrashort 800nm pulse~\cite{kaindl}. If $15\,$fs
long bandwidth-limited pulses centered at 800nm are used for this
purpose the generated mid-IR wavelength is about $10\,\mu$m at
which wavelength a 30fs long pulse would possess only one optical
cycle. The basic limitation in the conversion efficiency when
using GaSe as previously demonstrated for this
purpose~\cite{kaindl} was its strong TPA at 800nm.

\subsection{\label{subsec: dfg} Continuous-wave mid-IR DFG}
We report in this subsection the first DFG experiments
($5.5<\lambda_1<11.3\,\mu\text{m}$) ever performed with LIS, whose
partial results were already used to derive the accurate Sellmeier
equations of Subsection~\ref{subsec: sellmeier}. Continuous-wave
(cw) DFG is most suited for high-resolution spectroscopy in the
mid-IR range: when both DFG laser sources have linewidths
$<1\,$MHz the mid-IR radiation at $\lambda_1$ allows to perform
sub-Doppler spectroscopy of heavy atmospheric molecules with only
$\sim 1\,\mu$W of power. Until now, cw DFG of two near-IR lasers
for the generation of deep mid-IR light either used
AGS~\cite{canarelli,chen96,dlee} or GaSe~\cite{chen98}. We
demonstrate that LIS is an alternative promising candidate, since
we report cw parametric generation up to $11.3\,\mu$m with an
uncoated yellowish LIS sample of dimension 5x5x10 mm$^3$
($L=10\,$mm), cut in the X-Y plane for type-II ($eoe$) DFG at
($\theta=90^\circ,\varphi=42^{\circ}$).

Two tunable cw single-frequency Ti:sapphire lasers were used as
difference-frequency mixing sources. The laser beams, orthogonally
polarized, were collinearly focused onto the crystal with a
35-cm-focal length lens, yielding a nearly optimal $w_0=55\,\mu$m
waist for both pump wavelengths $\lambda_3<\lambda_2$. The pump
wavelengths were blocked after the sample by a 1-mm thick uncoated
germanium (Ge) filter and the mid-IR radiation was detected by a
calibrated liquid-N$_2$-cooled HgCdTe photoconductive detector
with a $1\times 1\,$mm$^2$ active area. The detected voltage was
amplified by a low-noise preamplifier and fed to a lock-in
amplifier.
  \begin{figure}[t]
    \includegraphics[width=7cm]{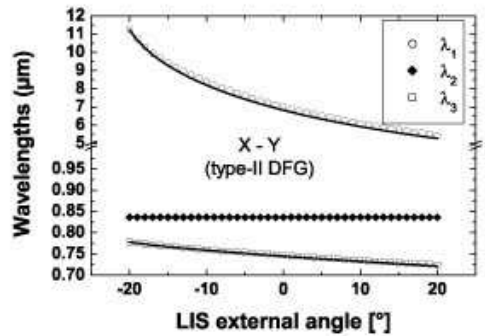}
    \caption{\label{fig: dfg-pm-11mic} Wavelength and angle tuning characteristics of type-II ($eoe$) DFG of two cw Ti:Saphire
    lasers in LIS. The solid lines are calculated from the Sellmeier equations of Subsection~\ref{subsec: sellmeier}.}
    \end{figure}

In Table~\ref{tab: table4}, we have already seen that at fixed
$\varphi=42^\circ$, wavelength tuning alone can cover only the
$6.6 - 7\,\mu$m range, limited by the tunability range of the
Ti:sapphire lasers (from curves 2-3 of Fig.~\ref{fig: opaxy}, a
tunability range extending into the blue would be required to
extend the DFG to $\sim 10\,\mu$m). Taking profit of angle tuning,
mid-IR radiation was generated with broad wavelength tunability in
the $5.5-11.3\,\mu$m spectral region with combined pump laser
wavelength ($\lambda_3$) and crystal angle ($\varphi$)
tuning~\cite{chen96}: in Fig.~\ref{fig: dfg-pm-11mic} one
Ti:sapphire laser was fixed at $\lambda_2\sim 836\,$nm, and the
other ($\lambda_3$) was tuned from 724 to 778 nm. The solid lines
are calculated phase-matching curves from Eqs.(\ref{eq:
sellmz1})-(\ref{eq: sellmz3}), highlighting the accuracy of the
Sellmeier equations that were fitted using the fixed-angle DFG
data of Table~\ref{tab: table4}.

\begin{figure}[t]
    \includegraphics[width=7cm]{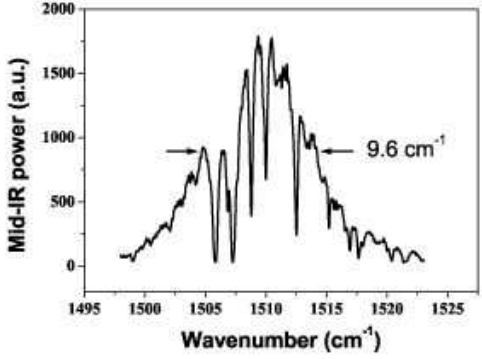}
    \caption{\label{fig: spectral-acceptance} Mid-IR DFG spectral tuning curve in pump tuning mode around
    $\lambda_3\simeq 765\,$nm. The other wavelength was fixed at
    $\lambda_2=865.3\,$nm.
     In this frequency scan, we could observe atmospheric water vapor absorption lines in laboratory ambient air
     over 42.5-cm open path between the LIS crystal and the HgCdTe detector.}
    \end{figure}

Fig.~\ref{fig: spectral-acceptance} displays the mid-IR spectral
tuning bandwidth around 1510 cm$^{-1}$ ($\lambda_1\sim
6.6\,\mu$m), when $\lambda_2$ is set at $865.3\,$nm and
$\lambda_3$ is tuned around $765$ nm, corresponding to a
quasi-normal incidence phase-matching. A $\Delta\bar{\nu}_1\sim
10\,$cm$^{-1}$ ($\bar{\nu}=1/\lambda\equiv \nu/c$) FWHM spectral
bandwidth is obtained, within which sharp atmospheric water vapor
absorption lines can be observed. The FWHM plane-wave spectral
acceptance for this pump tuning mode expresses as~\cite{dmitriev}
\begin{equation} \label{eq: spec-accept}
    \Delta \nu_1=1.722\frac{\pi}{L}
    \left[\frac{\partial\Delta k}{\partial
    \nu_1}\right]^{-1}_{\nu_1^{\text{PM}}}
\end{equation}
where $\partial \Delta k/\partial \nu_1$ can be expanded, taking
the constraint $\nu_3-\nu_2=\nu_1$ into account, as
\begin{equation} \label{eq: Dkdnu}
    \begin{split}
    \frac{\partial \Delta k}{\partial \nu_1}&=\frac{2\pi}{c}\left[n_3(\varphi)-\lambda_3 \frac{\partial n_3(\varphi)}{\partial \lambda_3}
    -n_1(\varphi)+\lambda_1 \frac{\partial n_1(\varphi)}{\partial
    \lambda_1}\right]\\
                                            &\equiv 2\pi\Delta_{31}.
    \end{split}
\end{equation}
In Eq.(\ref{eq: Dkdnu}), the angle-dependent indices are the
extraordinary indices in the X-Y plane. From (\ref{eq:
spec-accept})-(\ref{eq: Dkdnu}), the calculated spectral
acceptance is $\Delta\nu_1=214\,$GHz ($\Delta\bar{\nu}_1\simeq
7.2\,$cm$^{-1}$) corresponding to a pump tuning over
$\Delta\lambda_3=0.42\,$nm. The bandwidth broadening observed in
Fig.~\ref{fig: spectral-acceptance} is attributed to the focused
beam effect~\cite{zondy-dfg}.

Let us note that despite the strong mid-IR absorption at
$11.3\,\mu$m (Fig.~\ref{fig: ftir}), we could still detect $\sim
3\,$nW mid-IR power for input powers $P_3=200\,$mW and
$P_2=265\,$mW. This power is generated by the last exit layers of
material. For the same incident powers, the detected power at
$\lambda_1=7\,\mu$m was $P_1=33.5\,$nW. The net conversion
efficiency at this wavelength, corrected for all Fresnel optical
loss, amounted to $\Gamma=P_1/P_3P_2 \simeq 1.6\,\mu$W/W$^2$.
Using the general focused-beam theory for DFG~\cite{zondy-dfg}
that evaluates the focusing function $h$ for type-II DFG for the
given focusing parameters, walk-off value
($|\rho_{1,3}|=0.97^\circ$) and absorption loss ($\alpha_3\sim
0.15\,$cm$^{-1}$, $\alpha_2\sim 0.1\,$cm$^{-1}$, $\alpha_1\sim
0.05\,$cm$^{-1}$), we derive an effective nonlinear coefficient
$d_{\text{eff}}(7\,\mu\text{m})\simeq 5.3\,$pm/V. This value is
slightly lower than the calculated value using Eq.(\ref{eq:
deff-XY}) and the determined values of $d_{24}$ and $d_{15}$
(Eqs.(\ref{eq: d24-fem})-(\ref{eq: d31-fem})),
$d_{\text{eff}}\simeq 6.55\,$pm/V. The difference may arise from
the residual atmospheric water absorption lines in this region.

\section{\label{sec: damage} Laser-induced damage in L\lowercase{i}I\lowercase{n}S$_{2}$}

Illumination with various laser sources resulted in the following
photo-induced phenomena observed in LIS, whose magnitude depends
on wavelength and pulse duration: (i) micro-cracking (at
$\lambda=5\,\mu$m using femtosecond pulses from a free-electron
laser~\cite{knippels}); (ii) dendrite structure formation (at
$\lambda=1.079\,\mu$m using nanosecond pulses); (iii)
photo-induced absorption (PIA) as grey-tracks (at
$\lambda=0.8\,\mu$m using femtosecond pulses). Optical damage of
(i) and (ii) kinds is irreversible, while PIA damage is to a
certain extent reversible. It disappears at lower levels
spontaneously after several hours of storage at room temperature
or several minutes of illumation with $\lambda>500\,$nm. The PIA
was detected first in femtosecond OPA experiments with a 1.5mm
thick LIS sample pumped by a 1kHz Ti:sapphire regenerative
amplifier delivering pulses at 800nm with a duration of 200fs and
energy of $200\,\mu$J~\cite{rotermund}. No surface damage was
observed in these experiments up to a peak on-axis intensity of
$140\,$GW/cm$^2$. In principle as with other materials optical
damage in the femtosecond regime is expected to occur indirectly
as a result of filamentation and self-focussing through
higher-order nonlinear effects. These studies permitted to
estimate only an upper limit on the nonlinear absorption losses at
$800\,$nm which were independent of polarization and presumably
caused by inclusions. They could not be fitted by a TPA process
and the estimated upper limit of TPA is $\beta_{\text{TPA}}\sim
0.04\,$cm/GW at $800\,$nm~\cite{rotermund}. This is about 100
times lower than the $\beta_{\text{TPA}} =3.5-5.6\,$cm/GW
(depending on the polarization) fitted for a 0.3mm thick AGS
crystal under identical conditions, which is a manifestation of
the advantages of LIS related to its larger bandgap. An updated
value for the upper TPA limit that we derived for a different
sample, namely the LIS crystal used in Subsection~\ref{subsec:
opa}, is only slightly higher: $\beta_{\text{TPA}} =0.05\,$cm/GW.
Note that recent TPA measurements on KNbO$_3$ (which belongs to
the same class mm2) at $846\,$nm using a $5$-mm thick sample
phase-matched for cw blue light generation yielded a TPA magnitude
($\beta_{\text{TPA}} =3.2\,$cm/GW) of the same order as that for
AGS~\cite{ludlow}. In the experiment of Ref.~\cite{rotermund} the
gray tracks formation (leading to PIA) was attributed to
non-phase-matched SHG and residual blue light absorption, i.e.
this damage mechanism is only indirectly related to the high
intensity of the femtosecond pulses that resulted in higher
probability for unwanted SHG. Subsequently differential absorption
(before and after illumination) was measured that had a magnitude
reaching several cm$^{-1}$ and extended from $400\,$nm up to the
near-IR ($1-2\,\mu$m)~\cite{singapore}. Well pronounced maxima of
this absorption were observed at $430\,$nm and $650\,$nm. This
photochromic induced-absorption is similar to the so-called
Blue-Induced Infrared Absorption (BLIIRA) observed in SHG
experiments with KNbO$_3$, for which an increased near-IR
absorption coefficient (0.011 to $0.056\,$cm$^{-1}$ depending on
the sample and corresponding to 2.5 to 9 times the usual linear
absorption) at the fundamental wavelength of $860\,$nm was
detected with $100\,$mW of blue SH power ($30\,$kW/cm$^2$)
generated with a Ti:sapphire laser~\cite{mabuchi}. The BLIIRA
damage was also reported to be reversible, after a few hours of
relaxation at room temperature, and could be minimized by heating
the crystal or pumping at longer wavelengths~\cite{shiv}.

In order to understand the origin of the PIA in LIS, we proceeded
with a direct illumination of LIS samples at
$300\,<\lambda<\,450\,$nm using a 1 kW Xe lamp filtered through a
monochromator for $\sim 10\,$ minutes: The same coloration effect
was indeed obtained, with a maximum effect at 360 nm excitation
and vanishing at wavelengths shorter than $300\,$nm. Thus the PIA
observed in the OPA experiments is really caused by linear
absorption of non-phase matched blue SHG by point defects present
in the lattice. In line with the photo-induced signal in
absorption spectra (see Fig.3 in Ref.~\cite{singapore}), a
complicated system of lines appears in the Electron Spin Resonance
(ESR) spectrum. Analysis of the angular dependence allowed us to
reveal the superposition of two spectra related to different
paramagnetic defects. One of them has a spin $S=1/2$ and an
anisotropic $g$-factor with the following principal values of the
$\mathbf{g}$-tensor: $g_{1}=2.160$, $g_{2}=2.023$, $g_{3}=2.023$.
This spectrum is associated with a center of axial symmetry with
$g_{1}$, directed along the In-S bond and is related to sulfur
S$^{-}$ ion in a regular site. The second spectrum comprising a
broad line at $H\sim 3300\,$Gauss ($g=2.0024$) is associated with
Li$^{0}$.

The interpretation of these results is the following. At
short-wavelength excitation ($\lambda<450\,$nm) the following
recharge process occurs in LIS:
\begin{equation}
    \text{S}^{2-}+\text{Li}^{+}\rightarrow
    \text{S}^{-}+\text{Li}^{0}.
\end{equation}
Such excitation produces free charge carriers which are captured
in some deep traps and can be released at $\lambda>500\,$nm
illumination. We established that the gray track formation is
maximized for excitation near 360 nm but dropped by one order of
magnitude at 420 nm, which means that in OPO/OPA experiments it
can be reduced by using slightly longer pump wavelength
$\lambda_{3}$. Note that the same process also takes place at
x-ray irradiation, which is used for crystal orientation when
fabricating optical elements: some darkening occurs sometimes in
the bulk sample during this operation. The PIA magnitude was found
to be dependent on the LIS composition~\cite{oslo}: The effect is
minimum in the case of nearly stoichiometric composition. The PIA
was also found to be much more pronounced in LIS than in LISe.

\begin{figure}[t]
    \includegraphics[width=7cm]{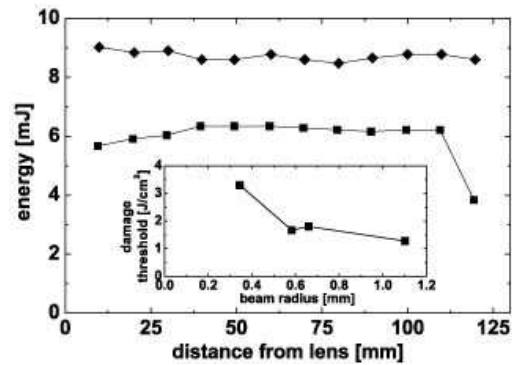}
    \caption{\label{fig: damage-setup} Measured pulse energy with crystal inserted (squares) and without it (diamonds).
    The drop in transmitted energy (last square)
    indicates damage formation. Inset: Damage threshold, expressed as fluence,
    versus beam radius at the crystal surface.}
  \end{figure}

Systematic optical damage threshold measurements were performed
with Nd:YAG lasers at 1064 nm - i.e. well within the clear
transmission range and far from the TPA region. Initial
measurements with a 6 W cw focused Nd:YAG laser beam failed to
produce measurable damage, and subsequent measurements were
performed with a pulsed laser. The pulse energy was varied between
6 mJ and 25 mJ at a repetition rate of 10 Hz and a pulse width of
about $10\,$ns. The collimated Gaussian beam with a beam radius of
$w_{in}=1.47\,$mm was focused with a 200 mm focal length lens to
produce a beam waist with a radius of $w_0$= $46\,\mu$m. The
crystal used was a rose sample with dimensions 4x5x5 mm$^3$
($\theta=90^{\circ},\varphi=24^{\circ}$). The corresponding
fluences ranged from $0.19$ J/cm$^2$ to $752$ J/cm$^2$, and by
using different pulse energies, the same fluence could be
sustained over a range of spot radii. The Gaussian character of
the beam was verified, and the beam radius measured by the
knife-edge technique. The onset of radiation damage was determined
from the observed sudden drop in the transmitted signal as shown
in Fig.~\ref{fig: damage-setup}. At each experimental point, the
crystal was inspected visually in a low-power microscope, and the
drop in transmission was found to correlate with the onset of
visible damage at the surface. By performing experiments with
different pulse energies it was found that the damage threshold,
expressed in J/cm$^2$, tends to decrease with increasing spot size
as seen in the inset figure of Fig.~\ref{fig: damage-setup}.
Although the amount of data is too limited to warrant a detailed
analysis, it suggests a levelling off at about 1 J/cm$^2$ at large
spot sizes (i.e. a peak intensity threshold of 100 MW/cm$^2$), and
this value agrees well with the value 1.1 J/cm$^2$ reported by
Knippels \emph{et al.} at a wavelength of 5 $\mu$m
\cite{knippels}. The observed dependence in the inset figure is
understandable in view of the fact that as the spot size
increases, it becomes increasingly difficult for heat generated at
the centre to diffuse to the unexposed regions of the crystal. In
Fig.~\ref{fig: damage-setup}, the initial increase of transmitted
energy reflects the fact that the center of the beam is too close
to the edge to allow the entire beam to be intercepted by the
crystal when the beam radius is largest. In some cases damage was
found to extend into the interior of the crystal.

\begin{table*}[ht]
\caption{\label{tab: table6}  Summary of the main physical
properties of LiInS$_2$.}
\begin{ruledtabular}
\begin{tabular}{ll}
\\
    \uppercase{crystallographic data} \\
    Structure & Wurtzite-type ($\beta$-NaFeO$_3$ structure)\\
    Symmetry, point group & Orthorhombic, mm2 (negative biaxial) \\
    Space group & Pna2$_1$ \\
    Lattice parameters (\text{\AA}) & $a=6.874$; $b=8.033$; $c=6.462$  (Ref.~\cite{isaenko-vasilyeva}) \\
    Principal axes assignment & $(X,Y,Z)\longleftrightarrow (b,a,c)$\\
    Density (g/cm$^3$) & 3.46 \\
    Microhardness (Pa) & 145$\times$10$^7$ (Refs.~\cite{kovach,kish84}) \\
\hline \\
    \uppercase{optical properties} \\
    Optical transmission ($\mu$m) & 0.34 - 11.5 \\
    Bandgap energy ($T=300$ K) (eV) & 3.55 ($E\parallel c$); 3.61 ($E\parallel b$); 3.58 ($E\parallel a$)  (Ref.~\cite{isaenko-vasilyeva}) \\
    Indices of refraction at $T=20^\circ$:  \\
    \multicolumn{1}{c}{$\lambda=0.532\mu$m} & $n_X=2.2413$; $n_Y=2.2824$; $n_Z=2.2991$ \\
    \multicolumn{1}{c}{$\lambda=1.064\mu$m} & $n_X=2.1288$; $n_Y=2.1645$; $n_Z=2.1734$ \\
    \multicolumn{1}{c}{$\lambda=10.6\mu$m}  & $n_X=2.0200$; $n_Y=2.0566$; $n_Z=2.0595$ \\
    Optical axis angles ($0.35\mu$m$<\lambda<11.5\mu$m) & $55^\circ<V_Z<80^{\circ}$\\
    Birefringence walk-off at 1064 nm (mrad) & $\rho_{X-Y}=14.7$;
    $\rho_{X-Z}=20$ \\
    SHG fundamental wavelength range (nm) & 1617 - 8710\\
    Thermo-optic coefficients (10$^{-5}/^\circ$C)\\
     \qquad at 1064 nm, $T=20^\circ$C       & $dn_X/dT=3.72$; $dn_Y/dT=4.55$; $dn_Z/dT=4.47$ \\
    Absorption coefficient at $1.064\,\mu$m (cm$^{-1}$) & $\alpha \leq 0.04$  \\
    Laser damage threshold: \\
    \multicolumn{1}{l}{\qquad $\bullet$ surface, at 10 ns, $f=10\,$Hz,\,$\lambda=1.064\,\mu$m}& \quad 1 J/cm$^{2}$ \\
    \multicolumn{1}{l}{\qquad $\bullet$ bulk, at 500 fs, $f=25\,$MHz,\,$\lambda=1.064\,\mu$m}& \quad $>6$ GW/cm$^2$ (Ref.~\cite{knippels})\\
    \multicolumn{1}{l}{\qquad $\bullet$ surface, at 200 fs, $f=1\,$kHz,\,$\lambda=0.8\,\mu$m}& \quad $>140$ GW/cm$^2$ (Ref.~\cite{rotermund})\\
    Two-photon absorption at $0.8\,\mu$m (cm/GW)& $\beta_{\text{TPA}}\leq 0.05$\\
    Nonlinear optical coefficients (pm/V) \\
    \multicolumn{1}{l}{\qquad $\bullet$ by cw SHG at $2.5\,\mu$m} & $d_{15}=7.9 (\pm 1.6)$; $d_{24}=5.7 (\pm 1.2)$ \\
    \multicolumn{1}{l}{\qquad $\bullet$ by femtosecond SHG at $2.3\,\mu$m} & $d_{31}=7.2 (\pm 0.4)$; $d_{24}=5.7 (\pm 0.6)$; $d_{33}=-16 (\pm 4)$ \\
\hline \\
    \uppercase{thermal properties} \\
    Melting point ($^\circ$C)& $T_{\text{melt}}\sim 1000$ \\
    Thermal expansion at $T=20^\circ$C (10$^{-5}/^\circ$C) & $\alpha_X=+1.64$; $\alpha_Y=+0.91$; $\alpha_Z=+0.68$\\
    Specific heat at $T=300\,$K (J/mol/K) & $C_p=92.9$ \\
    Thermal conductivity at $T=300\,$K (W/m/$^\circ$C) & $K_X=6.2$; $K_Y=6.0$; $K_Z=7.6$ (Ref.~\cite{ebbers})\\
\hline \\
    \uppercase{electrical properties} \\
    Pyroelectric coefficient at $T=300\,$K ($\mu$C/m$^2$K) &
    6 (Ref.~\cite{bidault})\\
    Pure electro-optic coefficients (pm/V) & $r_{13}=1.0$;
    $r_{23}=0.4$; $r_{33}=-1.3$ \\
    Piezo-electric coefficients (pm/V) & $\bar{d}_{31}=-5.6$;
    $\bar{d}_{32}=-2.7$; $\bar{d}_{33}=+7.5$ \\
    Electric conductivity ($\Omega^{-1}$cm$^{-1}$) & $\sigma_Z=4.3\times
    10^{-10}$ (Ref.~\cite{zelt}); $\sigma=1\times
    10^{-12}$ (Refs.~\cite{schumann,kish84})\\

\end{tabular}
\end{ruledtabular}
\end{table*}

These damage properties should be discussed in comparison with
available damage data on AGS or AGSe. For AGS under cw pumping the
onset of thermal lensing has been previously evaluated at $\sim
11\,$kW/cm$^2$ from a cw AGS doubly resonant OPO containing $\sim
3\,$W of circulating power at 1.26 and 2.52 $\mu$m focused to a
waist of $w_{0}=90\,\mu$m~\cite{douillet1,douillet2}. The cw
damage threshold value is actually unknown, but is estimated to
occur in the range $\sim 20-40\,$kW/cm$^2$~\cite{catella}, despite
a report that 3.5 W of cw Nd:YAG radiation focused to a waist of
$20\,\mu$m ($280\,$kW/cm$^{2}$) did not damage an AGS
sample~\cite{simon}. For AGSe, the measured damage threshold at
$9\,\mu$m (cw CO$_2$ laser) was ascertained to be
$5-22\,$kW/cm$^2$~\cite{acharekar}, corroborating that
chalcopyrites display extremely low cw damage threshold. The fact
that for 6W at 1064 nm, focused down to $w_{0}=40\,\mu$m no
visible damage was observed in the LIS sample sets the cw damage
threshold to a lower limit of $>120\,$kW/cm$^2$. The 5 times
larger thermal conductivity of LIS as compared to AGS certainly
favors the use of LIS in intra-cavity cw applications. Under
similar nanosecond pulsed regime, however, the $\sim
1\,$J/cm$^{2}$ damage threshold measured with LIS is comparable to
the threshold found for AGS or AGSe~\cite{kato97,ziegler}.
Independent measurements at longer wavelengths ($9.55\,\mu$m) with
$30\,$ns long pulses yielded, under identical experimental
conditions, damage thresholds of $6.5\,$J/cm$^2$ for
LIS~\cite{grechin} and $5.5\,$J/cm$^2$ and $4.2\,$ J/cm$^2$ for
AGS and AGSe, respectively~\cite{badikov}.

From these preliminary damage threshold investigations, it is
premature to assess whether the damage mechanisms are due to
native point defects or are intrinsic to the compound. Considering
the current state-of-the-art of residual absorption on this
compound, one may expect that with an improvement in residual
transparency and a better control of the growth parameters toward
the production of stoichiometric material, the above damage
threshold values can be improved.

\section{Conclusions}

We have characterized the main physical properties of the newly
emerging lithium thioindate chalcogenide semiconductor that are
relevant for nonlinear applications from the visible up to the
deep mid-IR. Table~\ref{tab: table6} summarizes these properties,
as well as those determined in previous works. LiInS$_{2}$ or LIS
can now be considered as a mature compound and its future
widespread use in nonlinear devices should refine all the data
collected from this investigation campaign. Among its most
relevant advantages over the existing mid-IR crystals, one may
quote its excellent thermal stability and high damage threshold
suitable for high power laser applications and its extremely
extended phase-matching capabilities (over its whole transparency
range) using either type-I or type-II interactions in or out-of
principal planes. Such an extended phase-matching capability is
still compatible with low birefringence walk-off angles. LIS is
the only existing material for direct down-conversion from the
near-IR to the deep mid-IR up to $12\,\mu$m, owing to its highest
bandgap energy. Its effective nonlinearity of $\sim 7\,$pm/V
(slightly lower than that of the chalcopyrite crystals) is
substantially higher than that of the oxides of the KTP family and
is almost constant for type-II interaction in the azimuthal
principal plane. We have shown that LIS should be very promising
for the frequency down-conversion of femtosecond laser sources. At
equivalent loss level, LIS would be less subject to thermal
lensing than AGS~\cite{douillet1,douillet2} permitting its use in
cw OPOs for deep mid-IR generation.

The next challenge and effort will be directed to the detailed
studies on its eventual ferro-electric properties. In contrast
with the chalcopyrites, LIS and LISe are pyroelectric as the
orthorhombic symmetry oxides (KTP isotypes) belonging to the same
$mm2$ point group. The spontaneous polarization is due to the fact
that in each bipyramid formed by two adjacent tetrapores
(Fig.~\ref{fig: lis-struc}) only one is occupied by a Li$^{+}$
cation. The other one is empty, which corresponds to a negative
effective charge. An intriguing issue, which has not still
received a definite answer, is whether such Li-chalcogenides can
be related to ferroelectric materials, i.e. if it is possible to
invert their spontaneous polarization along the polar $c$-axis. In
their short note~\cite{negran}, Negran \emph{et al} concluded that
LIS and LISe do not display a para-to-ferroelectric phase
transition from the absence of a sudden anomalous rise in the
record of the pyro-electric current versus temperature. This is
not surprising because of the screening effect of the high ionic
conductivity of Li. They measured a pyroelectric coefficient
$p^\sigma(300K)=2.6\times 10^{-10}$C/(cm$^{2}\cdot$K). A more
recent and reliable measurement of the pyro-electric coefficient
performed with our rose samples yielded a 2.3 times higher value
(Table~\ref{tab: table6}), which is $~5$ times smaller than the
value for KTP~\cite{pyroktp}. In contrast with Negran \emph{et al}
's report, a strong increase of the pyro-current slightly above
room-temperature, preceded by a sharp current spike, was noticed
but this increase was attributed to space charge relaxation of the
thermally-activated Li$^{+}$ cations~\cite{bidault}.

The issue about whether Li-chalcogenides are (or are not) related
to ferroelectrics is of upmost importance, considering their deep
mid-IR transparency which would allow to fabricate quasi-phase
matched periodically-poled structures for wavelengths longer than
$\sim 4-5\,\mu$m, which is the current limitation of oxide-based
ferroelectrics such as LiNbO$_{3}$ or KTiOPO$_3$. The measurement
of the Curie temperature (a temperature at which the material
tends to loose its spontaneous polarization), the study of its
dielectric conductivity versus an applied ac electric field in
order to reveal a dielectric hysteresis loop (and the value of the
coercive field) have to be performed to answer to the question.

Finally, we started to investigate other Li-containing compounds
such as LiInSe$_2$ (LISe), LiGaS$_2$ (LGS), LiGaSe$_2$ (LGSe),
LiGaTe$_2$ (LGTe)~\cite{isaenko-petrov,lga}.

\begin{acknowledgments} This Concerted Action could not have been
possible without the grant awarded by the European Commission
(DG12-MZCN directorate), in the frame of its INternational
CO-operation program of the 4th Framework Program (INCO-Copernicus
contract No IC15-CT98-0814). The authors are highly grateful to
J.F. Bardeau (Laboratoire de Physique de l'Etat Condens\'{e}, Le
Mans, France) and to P. Simon and his collaborators (Centre de
Recherche sur les Mat\'{e}riaux \`{a} Haute Temp\'{e}rature,
Orl\'{e}ans, France) for their expert collaboration in the Raman
and IR studies. We also acknowledge V. Drebushchak (Novosibirsk
State University, Russia) for the measurement of the specific
heat, V. Nadolinny (Institute of Inorganic Chemistry, SB RAS,
Novosibirsk) for the Electron Spin Resonance results and J.M.
Weulersse, M. Gilbert and G. Mennerat (Commissariat \`{a}
l'Energie Atomique, Saclay, France) for the
use of their nanosecond LiNbO$_3$-based OPO.\\
\end{acknowledgments}

\newpage 

%


\end{document}